\documentclass[prd,aps,floatfix,superscriptaddress,onecolumn]{revtex4-2}
\usepackage{amsfonts,amsmath,amssymb,amsthm}
\usepackage{bm}
\usepackage{booktabs}
\usepackage{braket} % Dirac bra-ket
\usepackage{cases}
\usepackage{color}
\usepackage{dcolumn}
\usepackage{epsfig}
\usepackage{epstopdf}   %{feynmp}
\usepackage{graphicx}
\usepackage[colorlinks,citecolor=blue,anchorcolor=red,menucolor=red,linkcolor=red,filecolor=red,runcolor=red,urlcolor=blue,frenchlinks=red]{hyperref}
\usepackage{indentfirst}
\usepackage{latexsym}
\usepackage{longtable}
\usepackage{mdframed}
\usepackage{multirow}
\usepackage{placeins}
\usepackage{rotating}
\usepackage{slashed}  %for Feynman symbols $\slashed{x}$
\usepackage{subfigure}
\usepackage{youngtab}
\newcommand{\etal}{\textit{et~al}.~}
\allowdisplaybreaks[4]
\begin{document}
%%%%%%%%%%%%%%%%%%%%%%%%%%%%%%%%%%%%%%%%%%%%%%%%%%%%%%%%%%%%%%%%%%%%
\title{Systematics of fully heavy dibaryons}
%%%%%%%%%%%%%%%%%%%%%%%%%%%%%%%%%%%%%%%%%%%%%%%%%%%%%%%%%%%%%%%%%%%%
\author{Xin-Zhen Weng}
\email{xinzhenweng@mail.tau.ac.il}
\affiliation{School of Physics and Astronomy, Tel Aviv University, Tel Aviv 6997801, Israel}
%%%
\author{Shi-Lin Zhu}
\email{zhusl@pku.edu.cn}
\affiliation{School of Physics and Center of High Energy Physics, Peking University, Beijing 100871, China}
%%%%%%%%%%%%%%%%%%%%%%%%%%%%%%%%%%%%%%%%%%%%%%%%%%%%%%%%%%%%%%%%%%%%
\date{\today}
%%%%%%%%%%%%%%%%%%%%%%%%%%%%%%%%%%%%%%%%%%%%%%%%%%%%%%%%%%%%%%%%%%%%
\begin{abstract}
%%%%%%%%%%%%%%%%%%%%%%%%%%%%%%%%%%%%%%%%%%%%%%%%%%%%%%%%%%%%%%%%%%%%

We systematically study the mass spectra of the fully heavy dibaryons in an extended chromomagnetic model, which includes both the colorelectric and chromomagnetic interactions.
We find no stable state below the corresponding baryon-baryon thresholds.
Besides the masses, we also estimate the relative width ratios of the two-body decay channels.
We hope our study will be of help for future experiments.
%

%%%%%%%%%%%%%%%%%%%%%%%%%%%%%%%%%%%%%%%%%%%%%%%%%%%%%%%%%%%%%%%%%%%%
\end{abstract}
%%%%%%%%%%%%%%%%%%%%%%%%%%%%%%%%%%%%%%%%%%%%%%%%%%%%%%%%%%%%%%%%%%%%

\maketitle
\thispagestyle{empty} % 首页不显示页码

%%%%%%%%%%%%%%%%%%%%%%%%%%%%%%%%%%%%%%%%%%%%%%%%%%%%%%%%%%%%%%%%%%%%
\section{Introduction}
\label{Sec:Introduction}
%%%%%%%%%%%%%%%%%%%%%%%%%%%%%%%%%%%%%%%%%%%%%%%%%%%%%%%%%%%%%%%%%%%%

In the past few decades, we have witnessed huge progress in searching for the exotic states, which cannot be explained as the conventional mesons or baryons.
In 2003, the Belle Collaboration observed the $X(3872)$ state in the exclusive $B^{\pm}{\to}K^{\pm}\pi^{+}\pi^{-}J/\psi$ decays~\cite{Belle:2003nnu}.
Its quantum numbers are $I^{G}J^{PC}=0^{+}1^{++}$~\cite{Zyla:2020zbs}.
Since then, lots of $XYZ$ states have been discovered, such as the
$Z_{c}(3900)$~\cite{BESIII:2013ris,Belle:2013yex},
$Y(3940)$~\cite{Belle:2004lle},
$Z_{c}(4020)$~\cite{BESIII:2013ouc},
$Y(4140)$~\cite{CDF:2009jgo},
$Y(4260)$~\cite{BaBar:2005hhc},
$Y(4360)$~\cite{BaBar:2006ait},
$Y(4660)$~\cite{Belle:2007umv},
$Z_{cs}(3985)$~\cite{BESIII:2020qkh}, $Z_{cs}(4000)$, and $Z_{cs}$(4020)~\cite{LHCb:2021uow},
$Z_{b}(10610)$, and $Z_{b}(10650)$~\cite{Bondar:2011aa}
and so on.
They are called the charmonium- or bottomonium-like states since they consist of at least a heavy $c\bar{c}$ or $b\bar{b}$ pair.
Besides the charmonium- and bottomonium-like states, various open heavy flavor exotic states have also been found in experiments.
In 2020, the LHCb Collaboration observed two singly-charmed resonances $X_{0/1}(2900)$ in the $D^{-}K^{+}$ channel~\cite{LHCb:2020bls,LHCb:2020pxc}.
Later, they observed a narrow doubly-charmed tetraquark $T_{cc}^{+}$ in the $D^{0}D^{0}\pi^{+}$ mass spectrum, just below the $D^{*+}D^{0}$ threshold~\cite{LHCb:2021vvq,LHCb:2021auc}.
Moreover, in 2020, the LHCb collaboration observed two structures in the $J/\psi$-pair invariant mass spectrum in the range of 6.2--7.2~GeV, which could be the $cc\bar{c}\bar{c}$ tetraquarks~\cite{LHCb:2020bwg}.
They are good candidates of the exotic structures like 
the compact tetraquark~\cite{Maiani:2004vq,Cui:2006mp,Ebert:2007rn,Park:2013fda,Anwar:2018sol},
the hybrid meson~\cite{Zhu:2005hp,Esposito:2016itg},
the molecule~\cite{Tornqvist:1993ng,Tornqvist:2004qy,Swanson:2003tb,Hanhart:2007yq,Carames:2010zz,Aceti:2012cb,Chen:2015add,Guo:2017jvc}, and so on.
More detailed reviews can be found in
Refs.~\cite{Chen:2016qju,Esposito:2016noz,Lebed:2016hpi,Ali:2017jda,Karliner:2017qhf,Olsen:2017bmm,Guo:2017jvc,Yuan:2018inv,Liu:2019zoy,Brambilla:2019esw,Chen:2022asf,Meng:2022ozq}.

Compared to the tetraquark systems, the experimental progress in the six-quark system is relatively scarce.
The first well-understood six-quark system is the deuteron observed by Urey {\etal} in 1932~\cite{Urey:1932gik,Urey:1932pvy}.
It is a loosely bound molecular state composed of a proton and a neutron, with a binding energy of only $2.2~\text{MeV}$.
Another dibaryon candidate is the $d^{*}(2380)$, with quantum numbers $IJ^{P}=03^{+}$, observed by the WASA-at-COSY Collaboration~\cite{Bashkanov:2008ih,WASA-at-COSY:2011bjg,WASA-at-COSY:2012seb,WASA-at-COSY:2014dmv,WASA-at-COSY:2014lmt,WASA-at-COSY:2014qkg}.
Theoretically, such a system was first explored by Dyson and Xuong in 1964~\cite{Dyson:1964xwa}, with an impressive prediction of its mass around $2350~\text{MeV}$.
In 1977, Kamae and Fujita used a non-relativistic one-boson-exchange (OBE) potential model to study the $\Delta\Delta$ system and found that the two $\Delta$ isobars are bound by about $100~\text{MeV}$~\cite{Kamae:1976at}.
More details can be found in the recent review Ref.~\cite{Clement:2020mab} and references therein.

Another interesting system would be the fully heavy dibaryon~\cite{Lyu:2021qsh,Mathur:2022nez,Liu:2021pdu,Alcaraz-Pelegrina:2022fsi,Richard:2020zxb,Huang:2020bmb}.
For such a system, the relativistic effects are negligible, and the kinetic energy is small since the constituent quarks are heavy.
In Ref.~\cite{Lyu:2021qsh}, Lyu~\etal used lattice QCD to study the $\Omega_{ccc}\Omega_{ccc}$ in the ${^{1}S_{0}}$ channel.
They found this system is loosely bound by about $5.68~\text{MeV}$.
Within the same methodology, Mathur~\etal replaced the charm quarks by the bottom quarks, and found a very deeply bound $\Omega_{bbb}\Omega_{bbb}$ dibaryon in the same channel, with a binding energy about $89~\text{MeV}$~\cite{Mathur:2022nez}.
A recent study in the extended OBE model also supported the existence of these bound states, although the binding energy for the $\Omega_{bbb}\Omega_{bbb}$ system is considerably smaller ($\sim6~\text{MeV}$)~\cite{Liu:2021pdu}.
On the other hand, Alcaraz-Pelegrina~\etal studied the fully heavy dibaryons with the Diffusion Monte Carlo method within quark model~\cite{Alcaraz-Pelegrina:2022fsi}.
They found that all these states are above the thresholds of the two fully heavy baryons.
In Ref.~\cite{Richard:2020zxb}, Richard~\etal explored the $bbbccc$ dibaryons and found no bound states below the lowest dissociation threshold as well.

For the fully heavy dibaryon, interactions are provided by gluon exchange and string confinement.
Usually, the interactions include the spin-independent Coulomb and confinement interactions, and the spin-dependent chromomagnetic, spin-orbit and tensor interactions~\cite{Eichten:1978tg,Isgur:1977ef,Godfrey:1985xj,Capstick:1986bm,DeRujula:1975qlm}.
When restricted to the ground state, the tensor and spin-orbit interactions can be ignored.
Then we have the simplified chromomagnetic model interactions
\begin{equation}
H_{\text{int}}
=
- \sum_{i<j} a_{ij} \bm{F}_{i}\cdot\bm{F}_{j} - \sum_{i<j} v_{ij}
\bm{S}_{i}\cdot\bm{S}_{j} \bm{F}_{i}\cdot\bm{F}_{j}\,.
\end{equation}
Note that for color-singlet hadrons the effective quark masses can be absorbed into the colorelectric interaction~\cite{Weng:2018mmf}.
These interactions give a good account of the ground state mesons and baryons~\cite{Weng:2018mmf}.
These effective interactions were used to study tetraquarks~\cite{Weng:2020jao,Guo:2021mja,Weng:2021hje,Guo:2021yws,Weng:2021ngd}, pentaquarks~\cite{Weng:2019ynv,An:2020vku,An:2020jix,An:2021vwi} and baryonia~\cite{Liu:2021gva}.
In this work, we use these interactions to study fully heavy dibaryons.
The paper is organized as follows.
In Sec.~\ref{Sec:Model}, we introduce the extended chromomagnetic model and construct the dibaryon wave functions.
We discuss the dibaryon masses and decay properties in Sec.~\ref{Sec:Result} and conclude in Sec.~\ref{Sec:Conclusion}.
%

%%%%%%%%%%%%%%%%%%%%%%%%%%%%%%%%%%%%%%%%%%%%%%%%%%%%%%%%%%%%%%%%%%%%
\section{The Extended Chromomagnetic Model}
\label{Sec:Model}
%%%%%%%%%%%%%%%%%%%%%%%%%%%%%%%%%%%%%%%%%%%%%%%%%%%%%%%%%%%%%%%%%%%%

In the quark model, the Hamiltonian of a $S$-wave hadron reads~\cite{Cui:2006mp,Buccella:2006fn,Luo:2017eub,Liu:2019zoy}
\begin{equation}\label{eqn:ECM}
H
%%%
=
\sum_{i}m_{i}+H_{\text{CE}}+H_{\text{CM}}\,,
\end{equation}
where $m_i$ is the effective mass of the $i$th quark (or antiquark).
$H_{\text{CE}}$ is the colorelectric (CE) interaction~\cite{Hogaasen:2013nca}
\begin{equation}\label{eqn:ECM:CE}
H_{\text{CE}}
%%%
=
-
\sum_{i<j}
a_{ij}
\bm{F}_{i}\cdot\bm{F}_{j}\,,
\end{equation}
and $H_{\text{CM}}$ is the chromomagnetic (CM) interaction
\begin{equation}\label{eqn:ECM:CM}
H_{\text{CM}}
%%%
=
-
\sum_{i<j}
v_{ij}
\bm{S}_{i}\cdot\bm{S}_{j}
\bm{F}_{i}\cdot\bm{F}_{j}\,.
\end{equation}
Here, $A_{ij}$ and $v_{ij}$ are the effective coupling constants which depend on the constituent quark masses and the spatial wave function.
$\bm{S}_{i}=\bm{\sigma}_i/2$ and $\bm{F}_{i}={\bm{\lambda}}_i/2$ are the quark spin and  color operators.
For the antiquark,
\begin{equation}
\bm{S}_{\bar{q}}=-\bm{S}_{q}^{*}\,,
\quad
\bm{F}_{\bar{q}}=-\bm{F}_{q}^{*}\,.
\end{equation}

Since
\begin{equation}\label{eqn:m+color=color}
\sum_{i<j}
\left(m_i+m_j\right)
\bm{F}_{i}\cdot\bm{F}_{j}
%%%
=
\left(\sum_{i}m_{i}\bm{F}_i\right)
\cdot
\left(\sum_{i}\bm{F}_{i}\right)
-
\frac{4}{3}
\sum_{i}
m_{i}\,,
\end{equation}
and the total color operator $\sum_i\bm{F}_i$ nullifies any color-singlet physical state, we can rewrite the Hamiltonian as~\cite{Weng:2018mmf}
\begin{equation}\label{eqn:hamiltonian:final}
H=
-\frac{3}{4}
\sum_{i<j}m_{ij}V^{\text{C}}_{ij}
-
\sum_{i<j}v_{ij}V^{\text{CM}}_{ij} \,,
\end{equation}
by introducing the quark pair mass parameter
\begin{equation}\label{eqn:para:color+m}
m_{ij}
=
\left(m_i+m_j\right)
+
\frac{4}{3}
a_{ij}\,,
\end{equation}
where $V^{\text{C}}_{ij}\equiv\bm{F}_{i}\cdot\bm{F}_{j}$ and $V^{\text{CM}}_{ij}\equiv\bm{S}_{i}\cdot\bm{S}_{j}\bm{F}_{i}\cdot\bm{F}_{j}$ are the colorelectric and CM interactions between quarks.
%
%To study the spectra of fully heavy dibaryons, we need to estimate the parameters $\{m_{ij},v_{ij}\}$.
Here $\{m_{ij},v_{ij}\}$ are unknown parameters.
In Ref.~\cite{Weng:2018mmf}, we fitted these parameters from the conventional mesons and baryons.
The related parameters are presented in Table~\ref{table:parameter:qq}.
In this work, we use the same set of parameters to study the ground state fully heavy dibaryons.
%
%%%%%%
%%%%%% parameter
%%%%%%
%
\begin{table*}[htbp]
\centering
\caption{Parameters of the $qq$ pairs for the baryons~\cite{Weng:2018mmf} (in units of $\text{MeV}$).}
\label{table:parameter:qq}
\begin{tabular}{lcccccccccccc}
\toprule[1pt]
\toprule[1pt]
%%%
Parameter&$m_{nn}^{b}$&$m_{ns}^{b}$&$m_{ss}^{b}$&$m_{nc}^{b}$&$m_{sc}^{b}$&$m_{cc}^{b}$&$m_{nb}^{b}$&$m_{sb}^{b}$&$m_{cb}^{b}$&$m_{b{b}}^{b}$\\
%%%
Value&$724.85$&$906.65$&$1049.36$&$2079.96$&$2183.68$&$3171.51$&$5412.25$&$5494.80$&$6416.07$&$9529.57$\\
%\midrule[1pt]
Parameter&$v_{n{n}}^{b}$&$v_{n{s}}^{b}$&$v_{ss}^{b}$&$v_{n{c}}^{b}$&$v_{s{c}}^{b}$&$v_{c{c}}^{b}$&$v_{n{b}}^{b}$&$v_{s{b}}^{b}$&$v_{c{b}}^{b}$&$v_{b{b}}^{b}$\\
%%%
Value&$305.34$&$212.75$&$195.30$&$62.81$&$70.63$&$56.75$&$19.92$&$8.47$&$31.45$&$30.65$\\
%%%
\bottomrule[1pt]
\bottomrule[1pt]
\end{tabular}
\end{table*}

To investigate the mass spectra of the dibaryon states, we need to construct the wave functions.
A detail construction of the dibaryon wave functions can be found in Appendix~\ref{App:WaveFunc}.
Diagonalizing the Hamiltonian in these bases, we can obtain the masses and eigenvectors of the fully heavy dibaryons.
%

%%%%%%%%%%%%%%%%%%%%%%%%%%%%%%%%%%%%%%%%%%%%%%%%%%%%%%%%%%%%%%%%%%%%
\section{Numerical results}
\label{Sec:Result}
%%%%%%%%%%%%%%%%%%%%%%%%%%%%%%%%%%%%%%%%%%%%%%%%%%%%%%%%%%%%%%%%%%%%

%%%%%%%%%%%%%%%%%%%%%%%%%%%%%%%%%%%%%%%%%%%%%%%%%%%%%%%%%%%%%%%%%%%%
\subsection{The ${c}^{6}$ and ${b}^{6}$ systems}
\label{Sec:c6+b6}
%%%%%%%%%%%%%%%%%%%%%%%%%%%%%%%%%%%%%%%%%%%%%%%%%%%%%%%%%%%%%%%%%%%%

\begin{figure*}[htbp]
\begin{tabular}{ccc}
%%%
%%%
\includegraphics[width=240pt]{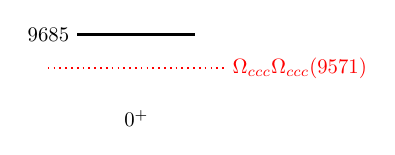}
&$\qquad$&
\includegraphics[width=240pt]{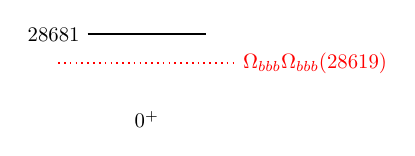}\\
%%%
(a) $cccccc$ states
&$\qquad$&
(b) $bbbbbb$ states\\
%%%
%%%
\includegraphics[width=240pt]{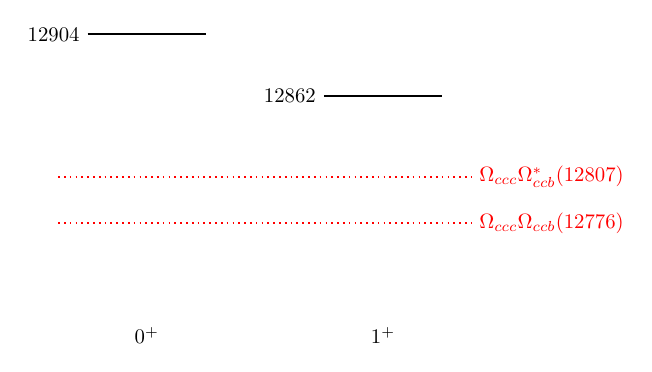}
&$\qquad$&
\includegraphics[width=240pt]{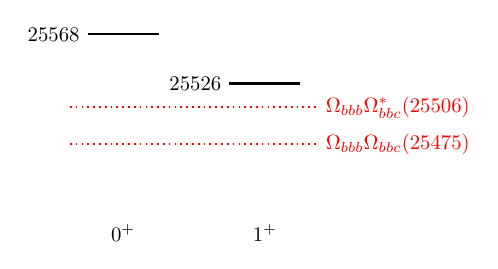}\\
%%%
(c) $cccccb$ states
&$\qquad$&
(d) $bbbbbc$ states\\
%%%
%%%
\includegraphics[width=240pt]{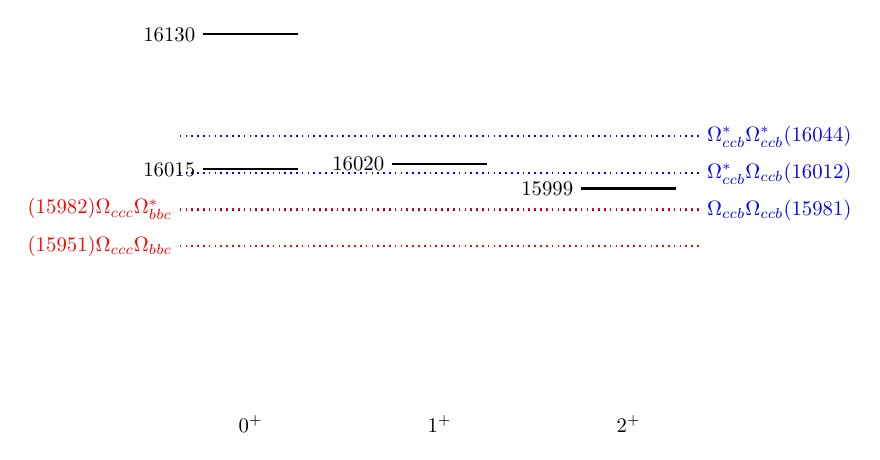}
&$\qquad$&
\includegraphics[width=240pt]{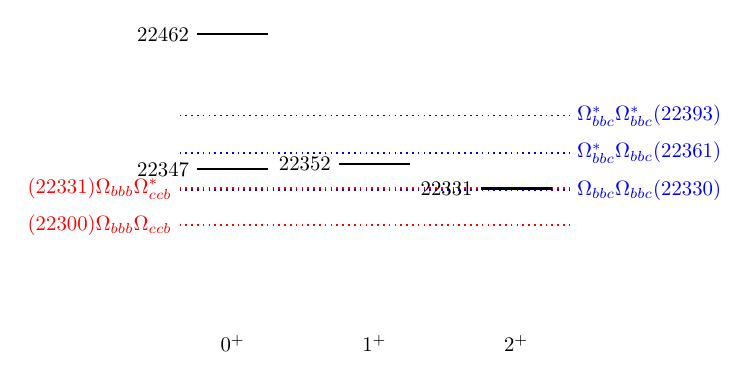}\\
%%%
(e) $ccccbb$ states
&$\qquad$&
(f) $bbbbcc$ states\\
%%%
%%%
\includegraphics[width=240pt]{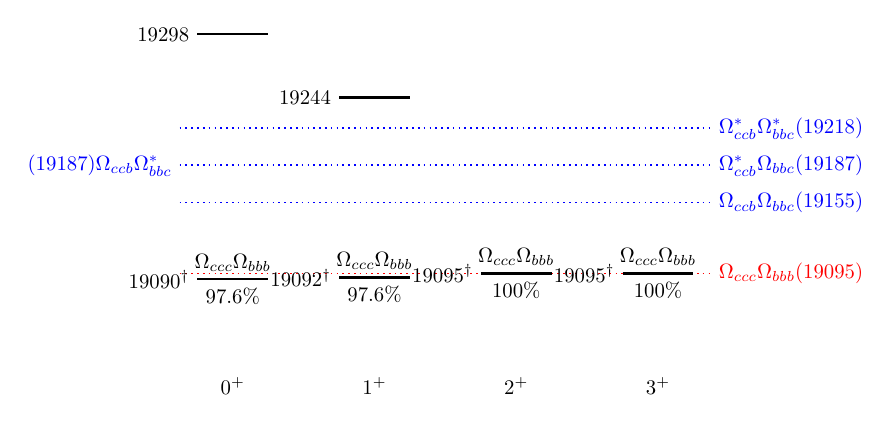}\\
%%%
(g) $cccbbb$ states\\
%%%
%%%
\end{tabular}
\caption{Masses of the fully heavy dibaryons. The dotted lines indicate various baryon-baryon thresholds, where the baryon masses are calculated in the same model~\cite{Weng:2018mmf}. The scattering states are marked with a dagger ($\dagger$), along with the proportion of their dominant components. The masses are all in units of MeV.}
\label{fig:6H}
\end{figure*}

Inserting the parameters into the Hamiltonian, we can obtain the mass spectra of dibaryons.
Their masses and eigenvectors are listed in Table~\ref{table:mass:Q6}.
In Fig.~\ref{fig:6H}, we plot their relative position along with baryon-baryon thresholds which they may decay into through quark rearrangement.
%
%%%
%%% mass:Q6
%%%
\begin{table}[htbp]
\centering
\caption{Masses and eigenvectors of the fully heavy dibaryons. All the masses are in units of MeV.}
\label{table:mass:Q6}
\begin{tabular}{ccccccccccc}
\toprule[1pt]
\toprule[1pt]
System&$J^{P}$&Mass&Eigenvector&Scattering~state\\
%%%
\midrule[1pt]
%%%
$cccccc$&$0^{+}$
&$9684.8$&$\{1\}$\\
%%%
%\midrule[1pt]
%%%
$bbbbbb$&$0^{+}$
&$28680.7$&$\{1\}$\\
%%%
\midrule[1pt]
%%%
$cccccb$&$0^{+}$
&$12904.0$&$\{1\}$\\
%%%
&$1^{+}$
&$12862.1$&$\{1\}$\\
%%%
%\midrule[1pt]
%%%
$bbbbbc$&$0^{+}$
&$25568.0$&$\{1\}$\\
%%%
&$1^{+}$
&$25526.0$&$\{1\}$\\
%%%
\midrule[1pt]
%%%
$ccccbb$
&$0^{+}$
&$16015.5$&$\{0.361,0.933\}$\\
&
&$16129.9$&$\{0.933,-0.361\}$\\
&$1^{+}$
&$16019.9$&$\{1\}$\\
&$2^{+}$
&$15999.0$&$\{1\}$\\
%%%
%\midrule[1pt]
%%%
$bbbbcc$
&$0^{+}$
&$22347.5$&$\{0.361,0.933\}$\\
&
&$22461.9$&$\{0.933,-0.361\}$\\
&$1^{+}$
&$22351.9$&$\{1\}$\\
&$2^{+}$
&$22330.9$&$\{1\}$\\
%%%
\midrule[1pt]
%%%
$cccbbb$&$0^{+}$
&$19090.4$&$\{-0.153,0.988\}$&$\Omega_{ccc}\Omega_{bbb}$\\
%%%
&
&$19297.9$&$\{0.988,0.153\}$\\
%%%
&$1^{+}$
&$19091.6$&$\{-0.155,0.988\}$&$\Omega_{ccc}\Omega_{bbb}$\\
%%%
&
&$19244.3$&$\{0.988,0.155\}$\\
%%%
&$2^{+}$
&$19095.3$&$\{1\}$&$\Omega_{ccc}\Omega_{bbb}$\\
%%%
&$3^{+}$
&$19095.3$&$\{1\}$&$\Omega_{ccc}\Omega_{bbb}$\\
%%%
\bottomrule[1pt]
\bottomrule[1pt]
\end{tabular}
\end{table}

First we consider the ${c}^{6}$ and ${b}^{6}$ systems.
Only the scalar states are allowed for these systems, namely the $D\left(cccccc,9684.8,0^{+}\right)$ and $D\left(bbbbbb,28680.7,0^{+}\right)$.
From Fig.~\ref{fig:6H}(a), we see that they are all above the baryon-baryon thresholds.
We may also see this from the Hamiltonian.
For the fully heavy dibaryon with the identical quarks, we have~\cite{Jaffe:1976ih,Park:2015nha}
\begin{equation}\label{eqn:ECM:6Q}
\Braket{H\left(Q^{6},0^{+}\right)}
=
\Braket{-\frac{3}{4}m_{QQ}^{D}\sum_{i<j}\bm{F}_{i}\cdot\bm{F}_{j}-v_{QQ}^{D}\sum_{i<j}\bm{F}_{i}\cdot\bm{F}_{j}\bm{S}_{i}\cdot\bm{S}_{j}}
=
3m_{QQ}^{D}+3v_{QQ}^{D}\,,
\end{equation}
where the superscript $D$ is an abbreviation of dibaryon.
On the other hand, the mass of the fully heavy $\Omega_{QQQ}$ baryon reads~\cite{Weng:2018mmf}
\begin{equation}
M_{\Omega_{QQQ}}
=
\frac{3}{2}m_{QQ}^{B}+\frac{1}{2}v_{QQ}^{B}\,,
\end{equation}
where the superscript $B$ is an abbreviation of the baryon.
In the present work, we assume that the dibaryons and the baryons share the same parameters, namely $m_{QQ}^{D}{\approx}m_{QQ}^{B}{\equiv}m_{QQ}$ and $v_{QQ}^{D}{\approx}v_{QQ}^{B}{\equiv}v_{QQ}$.
Thus the dibaryons are above the baryon-baryon thresholds by 
\begin{equation}
\Delta{E}
\approx
2v_{QQ}
=
\left\{
\begin{split}
113.5~\text{MeV}~\text{ for }~cccccc\,,\\
61.3~\text{MeV}~\text{ for }~bbbbbb\,.
\end{split}
\right.
\end{equation}
Of course, applying the baryon parameters to the dibaryon systems will cause some uncertainties.
Note that the dibaryon systems should have larger size compared to the baryon systems.
Thus the distance between two quarks within the dibaryons should be larger than that of the baryons.
So the attractive force between two quarks within the dibaryons should be weaker than that of the baryons.
Consequently, the realistic masses of the dibaryons should be slightly larger than the masses calculated in this work~\cite{An:2020vku,Deng:2020iqw}.
We see that even with this consideration, the $c^{6}$ ($b^{6}$) dibaryon should be above the $\Omega_{ccc}\Omega_{ccc}$ ($\Omega_{bbb}\Omega_{bbb}$) threshold.
%

%%%%%%%%%%%%%%%%%%%%%%%%%%%%%%%%%%%%%%%%%%%%%%%%%%%%%%%%%%%%%%%%%%%%
\subsection{The ${c}^{5}{b}$ and ${b}^{5}c$ systems}
\label{Sec:c5b+b5c}
%%%%%%%%%%%%%%%%%%%%%%%%%%%%%%%%%%%%%%%%%%%%%%%%%%%%%%%%%%%%%%%%%%%%

Next we consider the ${c}^{5}{b}$ and ${b}^{5}c$ systems.
In both cases, the lightest states have quantum numbers $J^{P}=1^{+}$.
They are $D(cccccb,12862.1,1^{+})$ and $D(bbbbbc,25526.0,1^{+})$, respectively.
From Fig.~\ref{fig:6H}, we can easily see that these states are all above thresholds.
Interestingly, numerical result suggests that the scalar state is heavier than the axial-vector one.
Let's consider the colorelectric interaction (here we use $cccccb$ as an example)
\begin{equation}\label{eqn:Hcolor:cccccb}
\Braket{H_{\text{C}}\left(cccccb\right)}
=
2m_{cc}+m_{cb}\,,
\end{equation}
which gives an identical contribution to the two states.
Thus the splitting comes from the chromomagnetic interaction.
By introducing the $\text{SU}(6)_{cs}=\text{SU}(3)_{c}{\otimes}\text{SU}(2)_{s}$ group~\cite{Jaffe:1976ih}, we have
\begin{equation}\label{eqn:HCM:cccccb}
\Braket{H_{\text{CM}}\left(cccccb\right)}
%%%
=
2v_{cc}+v_{cb}\left[1+\frac{S\left(S+1\right)}{12}-\frac{\Braket{\text{C}_{6}\left(cccccb\right)}}{32}\right]\,,
\end{equation}
where $\text{C}_{6}$ is the $\text{SU}(6)_{cs}$ Casimir operator.
The Pauli principle requires that the five charm quarks be anti-symmetric, thus we have two possible $\text{SU}(6)_{cs}$ representations
\begin{equation}\label{eqn:rep:SU6:cccccb}
\Yvcentermath1
\young(c,c,c,c,c,b)_{cs}
\sim
\young(cc,cc,cb)_{c}
\otimes
\young(ccc,ccb)_{s}\,,
%%%
\qquad
%%%
\young(cb,c,c,c,c)_{cs}
\sim
\young(cc,cc,cb)_{c}
\otimes
\young(cccb,cc)_{s}\,.
\end{equation}
The first one ($J=0$) is a $\text{SU}(6)_{cs}$ singlet with $\braket{\text{C}_{6}}=0$, while the second one ($J=1$) is a 35-plet with $\braket{\text{C}_{6}}=48$~\cite{Jaffe:1976ih}.
Thus we have
\begin{equation}\label{eqn:HCM:cccccb:2}
\Braket{H_{\text{CM}}\left(cccccb\right)}
%%%
=
\left\{
\begin{split}
&2v_{cc}+v_{cb}~\text{ for }~J=0\,,\\
&2v_{cc}-\frac{v_{cb}}{3}~\text{ for }~J=1\,.
\end{split}
\right.
\end{equation}
We conclude that the chromomagnetic interaction favors the axial-vector state.

Besides the spectra, the eigenvectors can also be used to estimate the decay properties of the dibaryons~\cite{Jaffe:1976ig,Strottman:1979qu,Zhao:2014qva,Wang:2015epa}.
We can calculate the overlap between the dibaryon and a particular baryon~$\times$~baryon channel.
Then we can estimate the decay amplitude of the dibaryon into that particular channel.
More precisely, we transform the wave function into the $QQQ{\otimes}QQQ$ configuration.
Usually, the $QQQ$ component in the dibaryon can be either of color-singlet or of color-octet.
The former one, $\ket{(QQQ)^{1_{c}}(QQQ)^{1_{c}}}$, can easily dissociate into two $S$-wave baryons in relative $S$ wave (the so-called ``Okubo-Zweig-Iizuka- (OZI-)superallowed'' decays), while the latter one, $\ket{(QQQ)^{8_{c}}(QQQ)^{8_{c}}}$, cannot fall apart without the gluon exchange.
For simplicity, we follow Refs.~\cite{Jaffe:1976ig,Jaffe:1976ih,Strottman:1979qu} and only consider the ``OZI-superallowed'' decays in this work.
In Table~\ref{table:decay:eigenvector}, we transform the dibaryon eigenvectors into the $QQQ{\otimes}QQQ$ configuration.
For simplicity, we only present the color-singlet components, and rewrite the bases as a direct product of two baryons
\begin{equation}
\Psi
=
\sum_{i}c_{i}\Ket{\psi_{i}\left(B{\otimes}B\right)}
+\cdots\,.
\end{equation}
For each decay channel, the decay width is proportional to the square of the coefficient $c_{i}$ of the corresponding component in the eigenvector, and also depends on the phase space.
For two body decay~\cite{Gao-1992-Group,Weng:2019ynv}
\begin{equation}\label{eqn:width}
\Gamma_{i}=\gamma_{i}\alpha\frac{k^{2L+1}}{m^{2L}}{\cdot}|c_i|^2,
\end{equation}
where $\gamma_{i}$ is a quantity related to the decay dynamics, $\alpha$ is an effective coupling constant, $k$ is the momentum of the final baryons in the rest frame of the initial dibaryon, $L$ is the relative partial wave between the two baryons, and $m$ is the dibaryon mass.
For the decay processes in this work, $(k/m)^2$'s are always of $\mathcal{O}(10^{-2})$ or even smaller.
Thus the higher wave decays are all suppressed.
We only consider the $S$-wave decays.
Next we have to estimate the $\gamma_{i}$.
Generally, $\gamma_{i}$ depends on the spatial wave functions of the initial and final states, which are different for each decay process.
In the quark model, the spatial wave functions of the ground state $1/2^{+}$ and $3/2^{+}$ baryons  are similar.
Thus for \emph{each} dibaryon, we have
\begin{equation}
\gamma_{B_{1}B_{2}}
=
\gamma_{B_{1}B_{2}^{*}}
=
\gamma_{B_{1}^{*}B_{2}}
=
\gamma_{B_{1}^{*}B_{2}^{*}}\,,
\end{equation}
where $B_{i}$ and $B_{i}^{*}$ denote the ground state $1/2^{+}$ and $3/2^{+}$ baryons with the same flavor contents.
In Tables~\ref{table:decay:kci2} and \ref{table:decay:Ratio}, we calculate the values of $k\cdot|c_i|^2$ and the relative widths for the fully heavy dibaryon decays.
The scalar state $D(cccccb,12904.0,0^{+})$ can easily decay into $\Omega_{ccc}\Omega_{ccb}^{*}$, but its phase space is quite small.
Thus it may not be very broad~\cite{Jaffe:1976ig}.
The axial-vector state $D(cccccb,12862.1,1^{+})$ can decay to the $\Omega_{ccc}\Omega_{ccb}^{(*)}$ channels with comparable widths.
More precisely,
\begin{equation}
\frac{\Gamma\left[D(cccccb,12862.1,1^{+}){\to}\Omega_{ccc}\Omega_{ccb}^{*}\right]}{\Gamma\left[D(cccccb,12862.1,1^{+}){\to}\Omega_{ccc}\Omega_{ccb}\right]}
\sim
1.004\,.
\end{equation}
The decay properties of the $bbbbbc$ dibaryon states are similar.
For example,
\begin{equation}
\frac{\Gamma\left[D(bbbbbc,25526.0,1^{+}){\to}\Omega_{bbb}\Omega_{bbc}^{*}\right]}{\Gamma\left[D(bbbbbc,25526.0,1^{+}){\to}\Omega_{bbb}\Omega_{bbc}\right]}
\sim
1.3\,.
\end{equation}
%
%%%
%%% decay:eigenvector
%%%
\begin{table*}[htbp]
\centering
\caption{The eigenvectors of the dibaryon states in various $QQQ{\otimes}QQQ$ configurations. The scattering states are marked with a dagger ($\dagger$). All the masses are in units of MeV.}
\label{table:decay:eigenvector}
\begin{tabular}{ccccccccccccccccccccccccccc}
\toprule[1pt]
\toprule[1pt]
%%%
&&&\multicolumn{2}{c}{$ccc{\otimes}ccb$}&&&&\multicolumn{2}{c}{$bbb{\otimes}bbc$}\\
\cmidrule(lr){4-5}
\cmidrule(lr){9-10}
%%%
System&$J^{P}$&Mass
&$\Omega_{ccc}\Omega_{ccb}^{*}$
&$\Omega_{ccc}\Omega_{ccb}$&
System&$J^{P}$&Mass
&$\Omega_{bbb}\Omega_{bbc}^{*}$
&$\Omega_{bbb}\Omega_{bbc}$\\
\midrule[1pt]
%%%
$cccccb$
&$0^{+}$
&$12904.0$&$-0.447$&&
$bbbbbc$&$0^{+}$
&$25568.0$&$-0.447$\\
%%%
&$1^{+}$
&$12862.1$&$-0.333$&$0.298$&
&$1^{+}$
&$25526.0$&$-0.333$&$0.298$\\
%%%
%%%
\midrule[1pt]
\midrule[1pt]
%%%
&&&\multicolumn{2}{c}{$ccc{\otimes}bbc$}&&\multicolumn{4}{c}{$ccb{\otimes}ccb$}\\
\cmidrule(lr){4-5}
\cmidrule(lr){7-10}
%%%
System&$J^{P}$&Mass
&$\Omega_{ccc}\Omega_{bbc}^{*}$
&$\Omega_{ccc}\Omega_{bbc}$&
&$\Omega_{ccb}^{*}\Omega_{ccb}^{*}$
&$\Omega_{ccb}^{*}\Omega_{ccb}$
&$\Omega_{ccb}\Omega_{ccb}^{*}$
&$\Omega_{ccb}\Omega_{ccb}$&\\
\midrule[1pt]
%%%
$ccccbb$
&$0^{+}$
&$16015.5$&$0.538$&&
&$0.009$&&&$-0.374$&\\
%%%
&
&$16129.9$&$-0.208$&&
&$-0.509$&&&$-0.213$&\\
%%%
&$1^{+}$
&$16019.9$&$0.527$&$0.236$&
&$\times$&$-0.236$&$-0.236$&$\times$&\\
%%%
&$2^{+}$
&$15999.0$&$0.408$&$0.408$&
&$-0.272$&$0.136$&$-0.136$&&\\
%%%
%%%
\midrule[1pt]
\midrule[1pt]
%%%
&&&\multicolumn{2}{c}{$bbb{\otimes}ccb$}&&\multicolumn{4}{c}{$bbc{\otimes}bbc$}\\
\cmidrule(lr){4-5}
\cmidrule(lr){7-10}
%%%
System&$J^{P}$&Mass
&$\Omega_{bbb}\Omega_{ccb}^{*}$
&$\Omega_{bbb}\Omega_{ccb}$&
&$\Omega_{bbc}^{*}\Omega_{bbc}^{*}$
&$\Omega_{bbc}^{*}\Omega_{bbc}$
&$\Omega_{bbc}\Omega_{bbc}^{*}$
&$\Omega_{bbc}\Omega_{bbc}$\\
\midrule[1pt]
%%%
$bbbbcc$
&$0^{+}$
&$22347.5$&$0.538$&&
&$0.009$&&&$-0.374$\\
%%%
&
&$22461.9$&$-0.208$&&
&$-0.509$&&&$-0.213$\\
%%%
&$1^{+}$
&$22351.9$&$0.527$&$0.236$&
&$\times$&$-0.236$&$-0.236$&$\times$\\
%%%
&$2^{+}$
&$22330.9$&$0.408$&$0.408$&
&$-0.272$&$0.136$&$-0.136$\\
%%%
%%%
\midrule[1pt]
\midrule[1pt]
%%%
&&&\multicolumn{4}{c}{$ccb{\otimes}bbc$}\\
\cmidrule(lr){4-7}
%%%
System&$J^{P}$&Mass
&$\Omega_{ccb}^{*}\Omega_{bbc}^{*}$
&$\Omega_{ccb}^{*}\Omega_{bbc}$
&$\Omega_{ccb}\Omega_{bbc}^{*}$
&$\Omega_{ccb}\Omega_{bbc}$\\
\midrule[1pt]
%%%
$cccbbb$&$0^{+}$
&$19090.4^{\dagger}$&$-0.042$&&&$-0.335$\\
&
&$19297.9$&$-0.456$&&&$0.107$\\
%%%
&$1^{+}$
&$19091.6^{\dagger}$&$0.015$&$-0.141$&$0.141$&$-0.272$\\
&
&$19244.3$&$-0.333$&$-0.172$&$0.172$&$0.222$\\
%%%
&$2^{+}$
&$19095.3^{\dagger}$&$0.111$&$-0.222$&$0.222$\\
%%%
&$3^{+}$
&$19095.3^{\dagger}$&$0.333$\\
%%%
\bottomrule[1pt]
\bottomrule[1pt]
\end{tabular}
\end{table*}
%
%%%
%%% decay:kci2
%%%
\begin{table*}[htbp]
\centering
\caption{The values of $k\cdot|c_{i}|^2$ for the dibaryon states (in units of MeV). The scattering states are marked with a dagger ($\dagger$).}
\label{table:decay:kci2}
\begin{tabular}{ccccccccccccccccccccccccccc}
\toprule[1pt]
\toprule[1pt]
%%%
&&&\multicolumn{2}{c}{$ccc{\otimes}ccb$}&&&&\multicolumn{2}{c}{$bbb{\otimes}bbc$}\\
\cmidrule(lr){4-5}
\cmidrule(lr){9-10}
%%%
System&$J^{P}$&Mass
&$\Omega_{ccc}\Omega_{ccb}^{*}$
&$\Omega_{ccc}\Omega_{ccb}$&
System&$J^{P}$&Mass
&$\Omega_{bbb}\Omega_{bbc}^{*}$
&$\Omega_{bbb}\Omega_{bbc}$\\
\midrule[1pt]
%%%
$cccccb$
&$0^{+}$
&$12904.0$&$152.6$&&
$bbbbbc$&$0^{+}$
&$25568.0$&$176.4$\\
%%%
&$1^{+}$
&$12862.1$&$63.7$&$64.0$&
&$1^{+}$
&$25526.0$&$55.6$&$71.4$\\
%%%
%%%
\midrule[1pt]
\midrule[1pt]
%%%
&&&\multicolumn{2}{c}{$ccc{\otimes}bbc$}&&\multicolumn{3}{c}{$ccb{\otimes}ccb$}\\
\cmidrule(lr){4-5}
\cmidrule(lr){7-9}
%%%
System&$J^{P}$&Mass
&$\Omega_{ccc}\Omega_{bbc}^{*}$
&$\Omega_{ccc}\Omega_{bbc}$&
&$\Omega_{ccb}^{*}\Omega_{ccb}^{*}$
&$\Omega_{ccb}^{*}\Omega_{ccb}$
&$\Omega_{ccb}\Omega_{ccb}$\\
\midrule[1pt]
%%%
$ccccbb$
&$0^{+}$
&$16015.5$&$137.3$&&
&$\times$&&$73.9$\\
%%%
&
&$16129.9$&$43.4$&&
&$216.0$&&$49.5$\\
%%%
&$1^{+}$
&$16019.9$&$140.1$&$37.9$&
&$\times$&$55.7$&$\times$\\
%%%
&$2^{+}$
&$15999.0$&$56.1$&$95.0$&
&$\times$&$\times$\\
%%%
%%%
\midrule[1pt]
\midrule[1pt]
%%%
&&&\multicolumn{2}{c}{$bbb{\otimes}ccb$}&&\multicolumn{3}{c}{$bbc{\otimes}bbc$}\\
\cmidrule(lr){4-5}
\cmidrule(lr){7-9}
%%%
System&$J^{P}$&Mass
&$\Omega_{ccc}\Omega_{bbc}^{*}$
&$\Omega_{ccc}\Omega_{bbc}$&
&$\Omega_{bbc}^{*}\Omega_{bbc}^{*}$
&$\Omega_{bbc}^{*}\Omega_{bbc}$
&$\Omega_{bbc}\Omega_{bbc}$\\
\midrule[1pt]
%%%
$bbbbcc$
&$0^{+}$
&$22347.5$&$117.6$&&
&$\times$&&$61.8$\\
%%%
&
&$22461.9$&$50.4$&&
&$228.0$&&$54.9$\\
%%%
&$1^{+}$
&$22351.9$&$127.4$&$40.6$&
&$\times$&$\times$&$\times$\\
%%%
&$2^{+}$
&$22330.9$&$\times$&$93.9$&
&$\times$&$\times$\\
%%%
%%%
\midrule[1pt]
\midrule[1pt]
%%%
&&&\multicolumn{4}{c}{$ccb{\otimes}bbc$}\\
\cmidrule(lr){4-7}
%%%
System&$J^{P}$&Mass
&$\Omega_{ccb}^{*}\Omega_{bbc}^{*}$
&$\Omega_{ccb}^{*}\Omega_{bbc}$
&$\Omega_{ccb}\Omega_{bbc}^{*}$
&$\Omega_{ccb}\Omega_{bbc}$\\
\midrule[1pt]
%%%
$cccbbb$&$0^{+}$
&$19090.4^{\dagger}$&$\times$&&&$\times$\\
&
&$19297.9$&$179.8$&&&$13.2$\\
%%%
&$1^{+}$
&$19091.6^{\dagger}$&$\times$&$\times$&$\times$&$\times$\\
&
&$19244.3$&$54.7$&$21.7$&$21.7$&$45.0$\\
%%%
&$2^{+}$
&$19095.3^{\dagger}$&$\times$&$\times$&$\times$\\
%%%
&$3^{+}$
&$19095.3^{\dagger}$&$\times$\\
%%%
\bottomrule[1pt]
\bottomrule[1pt]
\end{tabular}
\end{table*}
%
%%%
%%% decay:Ratio
%%%
\begin{table*}[htbp]
\centering
\caption{The partial width ratios for the dibaryon states. For each state, we choose one mode as the reference channel, and the partial width ratios of the other channels are calculated relative to this channel. The scattering states are marked with a dagger ($\dagger$). All the masses are in units of MeV.}
\label{table:decay:Ratio}
\begin{tabular}{ccccccccccccccccccccccccccc}
\toprule[1pt]
\toprule[1pt]
%%%
&&&\multicolumn{2}{c}{$ccc{\otimes}ccb$}&&&&\multicolumn{2}{c}{$bbb{\otimes}bbc$}\\
\cmidrule(lr){4-5}
\cmidrule(lr){9-10}
%%%
System&$J^{P}$&Mass
&$\Omega_{ccc}\Omega_{ccb}^{*}$
&$\Omega_{ccc}\Omega_{ccb}$&
System&$J^{P}$&Mass
&$\Omega_{bbb}\Omega_{bbc}^{*}$
&$\Omega_{bbb}\Omega_{bbc}$\\
\midrule[1pt]
%%%
$cccccb$
&$0^{+}$
&$12904.0$&$1$&&
$bbbbbc$&$0^{+}$
&$25568.0$&$1$\\
%%%
&$1^{+}$
&$12862.1$&$1$&$1.004$&
&$1^{+}$
&$25526.0$&$1$&$1.3$\\
%%%
%%%
\midrule[1pt]
\midrule[1pt]
%%%
&&&\multicolumn{2}{c}{$ccc{\otimes}bbc$}&&\multicolumn{3}{c}{$ccb{\otimes}ccb$}\\
\cmidrule(lr){4-5}
\cmidrule(lr){7-9}
%%%
System&$J^{P}$&Mass
&$\Omega_{ccc}\Omega_{bbc}^{*}$
&$\Omega_{ccc}\Omega_{bbc}$&
&$\Omega_{ccb}^{*}\Omega_{ccb}^{*}$
&$\Omega_{ccb}^{*}\Omega_{ccb}$
&$\Omega_{ccb}\Omega_{ccb}$\\
\midrule[1pt]
%%%
$ccccbb$
&$0^{+}$
&$16015.5$&$1$&&
&$\times$&&$1$~\footnote{For a dibaryon state decays into two channels through two different quark rearrangements, the $\gamma_{i}$'s, which depend on the wave functions of final states, may not equal (or approximately equal). Thus we compare only the decay width ratios between channels through same quark rearrangement type.}\\
%%%
&
&$16129.9$&$1$&&
&$4.4$&&$1$\\
%%%
&$1^{+}$
&$16019.9$&$1$&$0.3$&
&$\times$&$1$&$\times$\\
%%%
&$2^{+}$
&$15999.0$&$1$&$1.7$&
&$\times$&$\times$\\
%%%
%%%
\midrule[1pt]
\midrule[1pt]
%%%
&&&\multicolumn{2}{c}{$bbb{\otimes}ccb$}&&\multicolumn{3}{c}{$bbc{\otimes}bbc$}\\
\cmidrule(lr){4-5}
\cmidrule(lr){7-9}
%%%
System&$J^{P}$&Mass
&$\Omega_{bbb}\Omega_{ccb}^{*}$
&$\Omega_{bbb}\Omega_{ccb}$&
&$\Omega_{bbc}^{*}\Omega_{bbc}^{*}$
&$\Omega_{bbc}^{*}\Omega_{bbc}$
&$\Omega_{bbc}\Omega_{bbc}$\\
\midrule[1pt]
%%%
$bbbbcc$
&$0^{+}$
&$22347.5$&$1$&&
&$\times$&&$1$\\
%%%
&
&$22461.9$&$1$&&
&$4.1$&&$1$\\
%%%
&$1^{+}$
&$22351.9$&$1$&$0.3$&
&$\times$&$\times$&$\times$\\
%%%
&$2^{+}$
&$22330.9$&$\times$&$1$&
&$\times$&$\times$\\
%%%
%%%
\midrule[1pt]
\midrule[1pt]
%%%
&&&\multicolumn{4}{c}{$ccb{\otimes}bbc$}\\
\cmidrule(lr){4-7}
%%%
System&$J^{P}$&Mass
&$\Omega_{ccb}^{*}\Omega_{bbc}^{*}$
&$\Omega_{ccb}^{*}\Omega_{bbc}$
&$\Omega_{ccb}\Omega_{bbc}^{*}$
&$\Omega_{ccb}\Omega_{bbc}$\\
\midrule[1pt]
%%%
$cccbbb$&$0^{+}$
&$19090.4^{\dagger}$&$\times$&&&$\times$\\
&
&$19297.9$&$13.6$&&&$1$\\
%%%
&$1^{+}$
&$19091.6^{\dagger}$&$\times$&$\times$&$\times$&$\times$\\
&
&$19244.3$&$2.5$&$1$&$1$&$2.1$\\
%%%
&$2^{+}$
&$19095.3^{\dagger}$&$\times$&$\times$&$\times$\\
%%%
&$3^{+}$
&$19095.3^{\dagger}$&$\times$\\
%%%
\bottomrule[1pt]
\bottomrule[1pt]
\end{tabular}
\end{table*}
%

%%%%%%%%%%%%%%%%%%%%%%%%%%%%%%%%%%%%%%%%%%%%%%%%%%%%%%%%%%%%%%%%%%%%
\subsection{The ${c}^{4}{b}^{2}$ and ${b}^{4}c^{2}$ systems}
\label{Sec:c4b2+b4c2}
%%%%%%%%%%%%%%%%%%%%%%%%%%%%%%%%%%%%%%%%%%%%%%%%%%%%%%%%%%%%%%%%%%%%

Next we consider the ${c}^{4}{b}^{2}$ and ${b}^{4}c^{2}$ systems.
As shown in Fig.~\ref{fig:6H}(e--f), the lowest states of these two systems have quantum numbers $J^{P}=2^{+}$.
They are the $D(ccccbb,15999.0,2^{+})$ and $D(bbbbcc,22330.9,2^{+})$.
The highest states are $D(ccccbb,16129.9,0^{+})$ and $D(bbbbcc,22461.9,0^{+})$.
Their splittings are all about $130~\text{MeV}$.

Among these states, the $0^{+}$ states are of particular interests since they have two bases.
Taking $ccccbb$ as an example, their color configurations are $\ket{(cccc)^{\bar{6}_{c}}{\otimes}(bb)^{6_{c}}}$ and $\ket{(cccc)^{3_{c}}{\otimes}(bb)^{\bar{3}_{c}}}$ respectively.
For simplicity, we denote them as $\bar{6}_{c}{\otimes}6_{c}$ and $3_{c}{\otimes}\bar{3}_{c}$.
In Table~\ref{table:mass:Q6}, we present the eigenvectors of the $ccccbb$ dibaryon states.
We see that the lower mass state $D(ccccbb,16015.5,0^{+})$ is dominated by the $3_{c}{\otimes}\bar{3}_{c}$ component ($86.9\%$), while the higher one $D(ccccbb,15999.0,0^{+})$ is dominated by the $\bar{6}_{c}{\otimes}6_{c}$ component.
The reason is that the colorelectric interaction favors the color-triplet configuration.
More precisely,
\begin{equation}\label{eqn:HC:ccccbb}
\Braket{H_{\text{C}}\left(ccccbb\right)}
%%%
=
2m_{cc}+m_{bb}
+
{\delta}m_{cb}
\begin{pmatrix}
-5\\
&-2
\end{pmatrix}\,,
\end{equation}
where
\begin{equation}
{\delta}m_{cb}
=
\frac{m_{cc}+m_{bb}-2m_{cb}}{4}
=
-32.77~\text{MeV}\,.
\end{equation}
The contribution of the colorelectric interaction to the $3_{c}{\otimes}\bar{3}_{c}$ configuration is smaller than the $\bar{6}_{c}{\otimes}6_{c}$ one by nearly $100~\text{MeV}$.
The chromomagnetic interaction [${\delta}v_{cb}=(v_{cc}+v_{bb}-2v_{cb})/4$]
\begin{equation}\label{eqn:HCM:ccccbb}
\Braket{H_{\text{CM}}\left(ccccbb\right)}
%%%
=
2v_{cc}+v_{bb}
-
{\delta}v_{cb}
\begin{pmatrix}
3\\&10/3
\end{pmatrix}
-
\frac{v_{cb}}{32}
\Braket{\text{C}_{6}\left(ccccbb\right)}
%\frac{v_{cc}+v_{bb}-2v_{cb}}{32}
%\left(\frac{112}{3}-\frac{2}{3}\text{C}_{2}-\text{C}_{3}\right)_{b^{2}}
\end{equation}
will mix the two bases.
It is interesting to note that the off-diagonal terms and the difference of the diagonal terms of the Hamiltonian are symmetric over the $c$ and $b$ quarks.
%
%So the $bbbbcc$ system should possess same difference.
%
In Table~\ref{table:decay:eigenvector}, we find that the mixing between the two scalar $ccccbb$ states is almost identical with the mixing between the two scalar $bbbbcc$ states.
This can be explained from the Hamiltonian.
From Eqs.~(\ref{eqn:HC:ccccbb}--\ref{eqn:HCM:ccccbb}), we see that the off-diagonal term is suppressed by $v_{cb}\sim1/m_{c}m_{b}$, while the diagonal terms are both of $\mathcal{O}(m_{b})$.
There ratio are even more suppressed.
Numerically, we have 
\begin{equation}
\Braket{H\left(ccccbb\right)}
%%%
\propto
\begin{pmatrix}
1&-0.0024\\
&0.9947
\end{pmatrix}\,,
%%%
\qquad
%%%
\Braket{H\left(bbbbcc\right)}
%%%
\propto
\begin{pmatrix}
1&-0.0017\\
&0.9962
\end{pmatrix}\,.
\end{equation}
Note that a scale factor does not affect the eigenvectors.
The difference of the two matrices begins from the third digits after the decimal point.
So it is natural that the two Hamiltonians give nearly same eigenvectors.

Next we consider their decay properties.
Similar to previous cases, we find that all states are above thresholds.
For the highest state $D(ccccbb,16129.9,0^{+})$, we have (see Table~\ref{table:decay:Ratio})
\begin{equation}
\Gamma_{\Omega_{ccb}^{*}\Omega_{ccb}^{*}}:
\Gamma_{\Omega_{ccb}\Omega_{ccb}}
\sim
4.4\,.
\end{equation}
Thus the $\Omega_{ccb}^{*}\Omega_{ccb}^{*}$ mode is dominant.
For the $1^{+}$ and $2^{+}$ states, we have
\begin{equation}
\frac{\Gamma\left[D(ccccbb,16019.9,1^{+}){\to}\Omega_{ccc}\Omega_{bbc}\right]}{\Gamma\left[D(ccccbb,16019.9,1^{+}){\to}\Omega_{ccc}\Omega_{bbc}^{*}\right]}
=
0.3\,,
\end{equation}
and
\begin{equation}
\frac{\Gamma\left[D(ccccbb,15999.0,2^{+}){\to}\Omega_{ccc}\Omega_{bbc}\right]}{\Gamma\left[D(ccccbb,15999.0,2^{+}){\to}\Omega_{ccc}\Omega_{bbc}^{*}\right]}
=
1.7\,.
\end{equation}
The decay properties of the $bbbbcc$ dibaryon states are similar.
%

%%%%%%%%%%%%%%%%%%%%%%%%%%%%%%%%%%%%%%%%%%%%%%%%%%%%%%%%%%%%%%%%%%%%
\subsection{The ${c}^{3}{b}^{3}$ system}
\label{Sec:c3b3}
%%%%%%%%%%%%%%%%%%%%%%%%%%%%%%%%%%%%%%%%%%%%%%%%%%%%%%%%%%%%%%%%%%%%

Now we turn to the $cccbbb$ system.
From Table~\ref{table:mass:Q6}, we see that the lowest eigenstate
\begin{equation}
D(cccbbb,19090.4,0^{+})
=
0.988
\Omega_{ccc}\Omega_{bbb}
+
\cdots\,.
\end{equation}
This state couples very strongly to the $\Omega_{ccc}\Omega_{bbb}$ channel.
Thus it is likely to be very broad and is just part of the continuum.
Actually, this kind of eigenstate also exists in the calculation of the tetraquarks/pentaquarks, where the lower mass states couple very strongly with the $\text{meson}\otimes\text{meson/baryon}$ channels~\cite{Cui:2006mp,Hogaasen:2005jv,Weng:2019ynv,Weng:2020jao,Weng:2021hje,Weng:2021ngd}.
Recently, a diffusion Monte Carlo simulation within the dynamical quark model also suggested such a state which is probably \emph{two independent baryons close to each other and not a compact hexaquark}~\cite{Alcaraz-Pelegrina:2022fsi}.
Moreover, the states of $19091.6~\text{MeV}$ (with $J^{P}=1^{+}$), $19095.3~\text{MeV}$ (with $J^{P}=2^{+}$) and $19095.3~\text{MeV}$ (with $J^{P}=3^{+}$) also couple strongly to the $\Omega_{ccc}\Omega_{bbb}$ channel.
They are also scattering states.
For clarity, we indicate these states in the last column of Table~\ref{table:mass:Q6}.
As shown in Fig.~\ref{fig:6H}(g), all scattering states (mark with $\dagger$) lie close to the $\Omega_{ccc}\Omega_{bbb}$ threshold.

After identifying the scattering states, there are still one scalar and one axial-vector genuine dibaryon states.
They lie above all baryon-baryon thresholds.
The higher mass state $D(cccbbb,19297.9,0^{+})$ can decay into $\Omega_{ccb}^{*}\Omega_{bbc}^{*}$ and $\Omega_{ccb}\Omega_{bbc}$ channels with relative width ratio
\begin{equation}
\Gamma_{\Omega_{ccb}^{*}\Omega_{bbc}^{*}}:\Gamma_{\Omega_{ccb}\Omega_{bbc}}
\sim13.6\,.
\end{equation}
Thus the $\Omega_{ccb}^{*}\Omega_{bbc}^{*}$ mode dominates.
The other one, $D(cccbbb,19244.3,1^{+})$, decays into all $\Omega_{ccb}^{(*)}\Omega_{bbc}^{(*)}$ channels with comparable widths
\begin{equation}
\Gamma_{\Omega_{ccb}^{*}\Omega_{bbc}^{*}}:
\Gamma_{\Omega_{ccb}^{*}\Omega_{bbc}}:
\Gamma_{\Omega_{ccb}\Omega_{bbc}^{*}}:
\Gamma_{\Omega_{ccb}\Omega_{bbc}}
\sim
2.5:1:1:2.1\,.
\end{equation}
%

%%%%%%%%%%%%%%%%%%%%%%%%%%%%%%%%%%%%%%%%%%%%%%%%%%%%%%%%%%%%%%%%%%%%
\section{Conclusions}
\label{Sec:Conclusion}
%%%%%%%%%%%%%%%%%%%%%%%%%%%%%%%%%%%%%%%%%%%%%%%%%%%%%%%%%%%%%%%%%%%%

In this work, we have systematically studied the fully heavy dibaryons in an extended chromomagnetic model, which consists of effective color-electric and color-magnetic (chromomagnetic) interactions.
We find that there is no dibaryon state below the corresponding baryon-baryon thresholds.
Our numerical results suggest that the energy levels are mainly determined by the effective color-electric interaction.
For example, the $cccc$/$bb$ cluster in the $ccccbb$ dibaryons can be a color-triplet or a color-sextet.
The effective color-electric interaction splits the two configurations and makes the color-triplet configuration lighter than the color-sextet one by nearly $100~\text{MeV}$, resulting in a clear two-band structure.
The chromomagnetic interaction contributes small splittings for the two bands.
We find that the lightest state always has a higher spin.
The reason is that the Pauli principle imposes large restriction over the wave functions.
More precisely, the color and spin wave functions must be coupled in some particular form restricted by the $\text{SU}(6)_{cs}=\text{SU}(3)_{c}{\otimes}\text{SU}(2)_{s}$ symmetry.
Then the chromomagnetic interaction depends not only on the $\text{SU}(2)_{s}$ Casimir operator (spin), but also on the $\text{SU}(3)_{c}$ and $\text{SU}(6)_{cs}$ Casimir operators ($\text{C}_{3}$ and $\text{C}_{6}$).
The $\text{C}_{6}$ term has opposite effect compared to the spin term [for example, see Eq.~\eqref{eqn:HCM:cccccb}].
When the $\text{C}_{6}$ term prevails, the higher spin state becomes lighter.

With the eigenvectors obtained, we have also studied the decay properties of the dibaryons.
We hope that future experiments can search for these states.
%

%%%%%%%%%%%%%%%%%%%%%%%%%%%%%%%%%%%%%%%%%%%%%%%%%%%%%%%%%%%%%%%%%%%%
\section*{Acknowledgments}
%%%%%%%%%%%%%%%%%%%%%%%%%%%%%%%%%%%%%%%%%%%%%%%%%%%%%%%%%%%%%%%%%%%%

X.~Z.~W thanks Dr. Ranjit Nayak for proofreading the manuscript.
We are grateful to Professor Marek Karliner for helpful discussion.
This project was supported by the National Natural Science Foundation of China (NSFC) under Grant No.~11975033 and No.~12070131001; and the NSFC-ISF under Grant No.~3423/19.
%

%%%%%%%%%%%%%%%%%%%%%%%%%%%%%%%%%%%%%%%%%%%%%%%%%%%%%%%%%%%%%%%%%%%%
\begin{appendix}
%%%%%%%%%%%%%%%%%%%%%%%%%%%%%%%%%%%%%%%%%%%%%%%%%%%%%%%%%%%%%%%%%%%%

%%%%%%%%%%%%%%%%%%%%%%%%%%%%%%%%%%%%%%%%%%%%%%%%%%%%%%%%%%%%%%%%%%%%
\section{The wave function}
\label{App:WaveFunc}
%%%%%%%%%%%%%%%%%%%%%%%%%%%%%%%%%%%%%%%%%%%%%%%%%%%%%%%%%%%%%%%%%%%%

In this section, we construct the dibaryon wave functions.
In principle, the total wave function is a direct product of the spatial, flavor, color and spin wave functions.
In this work, we consider the ground state and assume that the spatial wave function is totally symmetric.
Then we need to construct the totally anti-symmetric $\text{flavor}\otimes\text{color}\otimes\text{spin}$ wave functions.
According to the flavor configurations, we can divide the fully heavy dibaryons into four categories, depending on their heavy flavor content:
\begin{enumerate}
\item $c^{6}$ and $b^{6}$,
\item $c^{5}b$ and $b^{5}c$,
\item $c^{4}b^{2}$ and $b^{4}c^{2}$,
\item $c^{3}b^{3}$.
\end{enumerate}
To satisfy the Pauli principle, we first construct the totally anti-symmetric wave functions for the clusters with the identical quarks (see Appendix~\ref{App:wavefunc:QN}), then we combine the clusters to construct the dibaryon wave functions.

For the $c^{6}$, $b^{6}$, $c^{5}b$ and $b^{5}c$ systems, we need to construct the wave functions in the $\left\{\left[\left(q_{1}q_{2}{\otimes}q_{3}\right){\otimes}q_{4}\right]{\otimes}q_{5}\right\}{\otimes}q_{6}$ configuration.
There are five possible color wave functions
\begin{align}
\phi_{\alpha1}
&{}=
\ket{(\{[(q_{1}q_{2})^{6}q_{3}]^{8}q_{4}\}^{\bar{6}}q_{5})^{\bar{3}}q_{6}}^{1}\,,
\notag\\
%%%
\phi_{\alpha2}
&{}=
\ket{(\{[(q_{1}q_{2})^{\bar{3}}q_{3}]^{8}q_{4}\}^{\bar{6}}q_{5})^{\bar{3}}q_{6}}^{1}\,,
\notag\\
%%%
\phi_{\alpha3}
&{}=
\ket{(\{[(q_{1}q_{2})^{6}q_{3}]^{8}q_{4}\}^{3}q_{5})^{\bar{3}}q_{6}}^{1}\,,
\notag\\
%%%
\phi_{\alpha4}
&{}=
\ket{(\{[(q_{1}q_{2})^{\bar{3}}q_{3}]^{8}q_{4}\}^{3}q_{5})^{\bar{3}}q_{6}}^{1}\,,
\notag\\
%%%
\phi_{\alpha5}
&{}=
\ket{(\{[(q_{1}q_{2})^{\bar{3}}q_{3}]^{1}q_{4}\}^{3}q_{5})^{\bar{3}}q_{6}}^{1}\,,
\end{align}
where the superscripts are color representations.
The spins of the dibaryons can be $0$, $1$, $2$ and $3$:
\begin{enumerate}
\item $J=0$:
\begin{align}
\chi^{0}_{\alpha1}
&{}=
\ket{(\{[(q_{1}q_{2})_{1}q_{3}]_{3/2}q_{4}\}_{1}q_{5})_{1/2}q_{6}}_{0}\,,
\notag\\
%%%
\chi^{0}_{\alpha2}
&{}=
\ket{(\{[(q_{1}q_{2})_{1}q_{3}]_{1/2}q_{4}\}_{1}q_{5})_{1/2}q_{6}}_{0}\,,
\notag\\
%%%
\chi^{0}_{\alpha3}
&{}=
\ket{(\{[(q_{1}q_{2})_{0}q_{3}]_{1/2}q_{4}\}_{1}q_{5})_{1/2}q_{6}}_{0}\,,
\notag\\
%%%
\chi^{0}_{\alpha4}
&{}=
\ket{(\{[(q_{1}q_{2})_{1}q_{3}]_{1/2}q_{4}\}_{0}q_{5})_{1/2}q_{6}}_{0}\,,
\notag\\
%%%
\chi^{0}_{\alpha5}
&{}=
\ket{(\{[(q_{1}q_{2})_{0}q_{3}]_{1/2}q_{4}\}_{0}q_{5})_{1/2}q_{6}}_{0}\,,
\end{align}
\item $J=1$:
\begin{align}
\chi^{1}_{\alpha1}
&{}=
\ket{(\{[(q_{1}q_{2})_{1}q_{3}]_{3/2}q_{4}\}_{2}q_{5})_{3/2}q_{6}}_{1}\,,
\notag\\
%%%
\chi^{1}_{\alpha2}
&{}=
\ket{(\{[(q_{1}q_{2})_{1}q_{3}]_{3/2}q_{4}\}_{1}q_{5})_{3/2}q_{6}}_{1}\,,
\notag\\
%%%
\chi^{1}_{\alpha3}
&{}=
\ket{(\{[(q_{1}q_{2})_{1}q_{3}]_{1/2}q_{4}\}_{1}q_{5})_{3/2}q_{6}}_{1}\,,
\notag\\
%%%
\chi^{1}_{\alpha4}
&{}=
\ket{(\{[(q_{1}q_{2})_{0}q_{3}]_{1/2}q_{4}\}_{1}q_{5})_{3/2}q_{6}}_{1}\,,
\notag\\
%%%
\chi^{1}_{\alpha5}
&{}=
\ket{(\{[(q_{1}q_{2})_{1}q_{3}]_{3/2}q_{4}\}_{1}q_{5})_{1/2}q_{6}}_{1}\,,
\notag\\
%%%
\chi^{1}_{\alpha6}
&{}=
\ket{(\{[(q_{1}q_{2})_{1}q_{3}]_{1/2}q_{4}\}_{1}q_{5})_{1/2}q_{6}}_{1}\,,
\notag\\
%%%
\chi^{1}_{\alpha7}
&{}=
\ket{(\{[(q_{1}q_{2})_{0}q_{3}]_{1/2}q_{4}\}_{1}q_{5})_{1/2}q_{6}}_{1}\,,
\notag\\
%%%
\chi^{1}_{\alpha8}
&{}=
\ket{(\{[(q_{1}q_{2})_{1}q_{3}]_{1/2}q_{4}\}_{0}q_{5})_{1/2}q_{6}}_{1}\,,
\notag\\
%%%
\chi^{1}_{\alpha9}
&{}=
\ket{(\{[(q_{1}q_{2})_{0}q_{3}]_{1/2}q_{4}\}_{0}q_{5})_{1/2}q_{6}}_{1}\,,
\end{align}
\end{enumerate}
where the subscripts are spins.
Note that $J=2$ and $J=3$ cases cannot satisfy the Pauli principle.

Combining the flavor, color and spin wave functions, we have the following bases
\begin{enumerate}
\item Type A: $\varphi_{\text{A}}=\{c^{6},b^{6}\}$
\begin{enumerate}
\item $J=0$:
\begin{equation}
\Psi_{\text{A}1}^{0}
=
\varphi_{\text{A}}
\otimes
\frac{1}{\sqrt{5}}
\left(
\phi_{\alpha1}
\chi^{0}_{\alpha5}
-
\phi_{\alpha2}
\chi^{0}_{\alpha4}
-
\phi_{\alpha3}
\chi^{0}_{\alpha3}
+
\phi_{\alpha4}
\chi^{0}_{\alpha2}
-
\phi_{\alpha5}
\chi^{0}_{\alpha1}
\right)\,,
\end{equation}
\end{enumerate}
\item Type B: $\varphi_{\text{B}}=\{c^{5}b,b^{5}c\}$
\begin{enumerate}
\item $J=0$:
\begin{equation}
\Psi_{\text{B}1}^{0}
=
\varphi_{\text{B}}
\otimes
\frac{1}{\sqrt{5}}
\left(
\phi_{\alpha1}
\chi^{0}_{\alpha5}
-
\phi_{\alpha2}
\chi^{0}_{\alpha4}
-
\phi_{\alpha3}
\chi^{0}_{\alpha3}
+
\phi_{\alpha4}
\chi^{0}_{\alpha2}
-
\phi_{\alpha5}
\chi^{0}_{\alpha1}
\right)\,,
\end{equation}
\item $J=1$:
\begin{equation}
\Psi_{\text{B}1}^{1}
=
\varphi_{\text{B}}
\otimes
\frac{1}{\sqrt{5}}
\left(
\phi_{\alpha1}
\chi^{1}_{\alpha9}
-
\phi_{\alpha2}
\chi^{1}_{\alpha8}
-
\phi_{\alpha3}
\chi^{1}_{\alpha7}
+
\phi_{\alpha4}
\chi^{1}_{\alpha6}
-
\phi_{\alpha5}
\chi^{1}_{\alpha5}
\right)\,.
\end{equation}
\end{enumerate}
\end{enumerate}

Next we consider the $c^{4}b^{2}$ and $b^{4}c^{2}$ systems.
Here we use the wave functions in the $\left[\left(q_{1}q_{2}{\otimes}q_{3}\right){\otimes}q_{4}\right]{\otimes}q_{5}q_{6}$ configuration.
The color wave functions read
\begin{align}
\phi_{\beta1}
&{}=
\ket{\{[(q_{1}q_{2})^{6}q_{3}]^{8}q_{4}\}^{\bar{6}}(q_{5}q_{6})^{6}}^{1}\,,
\notag\\
%%%
\phi_{\beta2}
&{}=
\ket{\{[(q_{1}q_{2})^{\bar{3}}q_{3}]^{8}q_{4}\}^{\bar{6}}(q_{5}q_{6})^{6}}^{1}\,,
\notag\\
%%%
\phi_{\beta3}
&{}=
\ket{\{[(q_{1}q_{2})^{6}q_{3}]^{8}q_{4}\}^{3}(q_{5}q_{6})^{\bar{3}}}^{1}\,,
\notag\\
%%%
\phi_{\beta4}
&{}=
\ket{\{[(q_{1}q_{2})^{\bar{3}}q_{3}]^{8}q_{4}\}^{3}(q_{5}q_{6})^{\bar{3}}}^{1}\,,
\notag\\
%%%
\phi_{\beta5}
&{}=
\ket{\{[(q_{1}q_{2})^{\bar{3}}q_{3}]^{1}q_{4}\}^{3}(q_{5}q_{6})^{\bar{3}}}^{1}\,.
\end{align}
And the spin wave functions read (the $J=3$ case does not satisfy the Pauli principle)
\begin{enumerate}
\item $J=0$:
\begin{align}
\chi^{0}_{\beta1}
&{}=
\ket{\{[(q_{1}q_{2})_{1}q_{3}]_{3/2}q_{4}\}_{1}(q_{5}q_{6})_{1}}_{0}\,,
\notag\\
%%%
\chi^{0}_{\beta2}
&{}=
\ket{\{[(q_{1}q_{2})_{1}q_{3}]_{1/2}q_{4}\}_{1}(q_{5}q_{6})_{1}}_{0}\,,
\notag\\
%%%
\chi^{0}_{\beta3}
&{}=
\ket{\{[(q_{1}q_{2})_{0}q_{3}]_{1/2}q_{4}\}_{1}(q_{5}q_{6})_{1}}_{0}\,,
\notag\\
%%%
\chi^{0}_{\beta4}
&{}=
\ket{\{[(q_{1}q_{2})_{1}q_{3}]_{1/2}q_{4}\}_{0}(q_{5}q_{6})_{0}}_{0}\,,
\notag\\
%%%
\chi^{0}_{\beta5}
&{}=
\ket{\{[(q_{1}q_{2})_{0}q_{3}]_{1/2}q_{4}\}_{0}(q_{5}q_{6})_{0}}_{0}\,,
\end{align}
\item $J=1$:
\begin{align}
\chi^{1}_{\beta1}
&{}=
\ket{\{[(q_{1}q_{2})_{1}q_{3}]_{3/2}q_{4}\}_{2}(q_{5}q_{6})_{1}}_{1}\,,
\notag\\
%%%
\chi^{1}_{\beta2}
&{}=
\ket{\{[(q_{1}q_{2})_{1}q_{3}]_{3/2}q_{4}\}_{1}(q_{5}q_{6})_{1}}_{1}\,,
\notag\\
%%%
\chi^{1}_{\beta3}
&{}=
\ket{\{[(q_{1}q_{2})_{1}q_{3}]_{1/2}q_{4}\}_{1}(q_{5}q_{6})_{1}}_{1}\,,
\notag\\
%%%
\chi^{1}_{\beta4}
&{}=
\ket{\{[(q_{1}q_{2})_{0}q_{3}]_{1/2}q_{4}\}_{1}(q_{5}q_{6})_{1}}_{1}\,,
\notag\\
%%%
\chi^{1}_{\beta5}
&{}=
\ket{\{[(q_{1}q_{2})_{1}q_{3}]_{3/2}q_{4}\}_{1}(q_{5}q_{6})_{0}}_{1}\,,
\notag\\
%%%
\chi^{1}_{\beta6}
&{}=
\ket{\{[(q_{1}q_{2})_{1}q_{3}]_{1/2}q_{4}\}_{1}(q_{5}q_{6})_{0}}_{1}\,,
\notag\\
%%%
\chi^{1}_{\beta7}
&{}=
\ket{\{[(q_{1}q_{2})_{0}q_{3}]_{1/2}q_{4}\}_{1}(q_{5}q_{6})_{0}}_{1}\,,
\notag\\
%%%
\chi^{1}_{\beta8}
&{}=
\ket{\{[(q_{1}q_{2})_{1}q_{3}]_{1/2}q_{4}\}_{0}(q_{5}q_{6})_{1}}_{1}\,,
\notag\\
%%%
\chi^{1}_{\beta9}
&{}=
\ket{\{[(q_{1}q_{2})_{0}q_{3}]_{1/2}q_{4}\}_{0}(q_{5}q_{6})_{1}}_{1}\,,
\end{align}
\item $J=2$:
\begin{align}
\chi^{2}_{\beta1}
&{}=
\ket{\{[(q_{1}q_{2})_{1}q_{3}]_{3/2}q_{4}\}_{2}(q_{5}q_{6})_{1}}_{2}\,,
\notag\\
%%%
\chi^{2}_{\beta2}
&{}=
\ket{\{[(q_{1}q_{2})_{1}q_{3}]_{3/2}q_{4}\}_{2}(q_{5}q_{6})_{0}}_{2}\,,
\notag\\
%%%
\chi^{2}_{\beta3}
&{}=
\ket{\{[(q_{1}q_{2})_{1}q_{3}]_{3/2}q_{4}\}_{1}(q_{5}q_{6})_{1}}_{2}\,,
\notag\\
%%%
\chi^{2}_{\beta4}
&{}=
\ket{\{[(q_{1}q_{2})_{1}q_{3}]_{1/2}q_{4}\}_{1}(q_{5}q_{6})_{1}}_{2}\,,
\notag\\
%%%
\chi^{2}_{\beta5}
&{}=
\ket{\{[(q_{1}q_{2})_{0}q_{3}]_{1/2}q_{4}\}_{1}(q_{5}q_{6})_{1}}_{2}\,.
\end{align}
\end{enumerate}
Then the possible total wave functions are
\begin{enumerate}
\setcounter{enumi}{2}
\item Type C: $\varphi_{\text{C}}=\{c^{4}b^{2},b^{4}c^{2}\}$
\begin{enumerate}
\item $J=0$:
\begin{align}
\Psi_{\text{C}1}^{0}
={}&
\varphi_{\text{C}}
\otimes
\frac{1}{\sqrt{2}}
\left(
\phi_{\beta1}
\chi^{0}_{\beta5}
-
\phi_{\beta2}
\chi^{0}_{\beta4}
\right)\,,
\notag\\
%%%
\Psi_{\text{C}2}^{0}
={}&
\varphi_{\text{C}}
\otimes
\frac{1}{\sqrt{3}}
\left(
\phi_{\beta3}
\chi^{0}_{\beta3}
-
\phi_{\beta4}
\chi^{0}_{\beta2}
+
\phi_{\beta5}
\chi^{0}_{\beta1}
\right)\,,
\end{align}
\item $J=1$:
\begin{equation}
\Psi_{\text{C}1}^{1}
=
\varphi_{\text{C}}
\otimes
\frac{1}{\sqrt{3}}
\left(
\phi_{\beta3}
\chi^{1}_{\beta4}
-
\phi_{\beta4}
\chi^{1}_{\beta3}
+
\phi_{\beta5}
\chi^{1}_{\beta2}
\right)\,,
\end{equation}
\item $J=2$:
\begin{equation}
\Psi_{\text{C}1}^{2}
=
\varphi_{\text{C}}
\otimes
\frac{1}{\sqrt{3}}
\left(
\phi_{\beta3}
\chi^{2}_{\beta5}
-
\phi_{\beta4}
\chi^{2}_{\beta4}
+
\phi_{\beta5}
\chi^{2}_{\beta3}
\right)\,,
\end{equation}
\end{enumerate}
\end{enumerate}

Finally, we construct the $c^{3}b^{3}$ bases in the $\left(q_{1}q_{2}{\otimes}q_{3}\right){\otimes}\left(q_{4}q_{5}{\otimes}q_{6}\right)$ configurations.
The color wave functions are
\begin{align}
\phi_{\gamma1}
&{}=
\ket{[(q_{1}q_{2})^{6}q_{3}]^{8}[(q_{4}q_{5})^{6}q_{6}]^{8}}^{1}\,,
\notag\\
%%%
\phi_{\gamma2}
&{}=
\ket{[(q_{1}q_{2})^{6}q_{3}]^{8}[(q_{4}q_{5})^{\bar{3}}q_{6}]^{8}}^{1}\,,
\notag\\
%%%
\phi_{\gamma3}
&{}=
\ket{[(q_{1}q_{2})^{\bar{3}}q_{3}]^{8}[(q_{4}q_{5})^{6}q_{6}]^{8}}^{1}\,,
\notag\\
%%%
\phi_{\gamma4}
&{}=
\ket{[(q_{1}q_{2})^{\bar{3}}q_{3}]^{8}[(q_{4}q_{5})^{\bar{3}}q_{6}]^{8}}^{1}\,,
\notag\\
%%%
\phi_{\gamma5}
&{}=
\ket{[(q_{1}q_{2})^{\bar{3}}q_{3}]^{1}[(q_{4}q_{5})^{\bar{3}}q_{6}]^{1}}^{1}\,.
\end{align}
And the spin wave functions are
\begin{enumerate}
\item $J=0$:
\begin{align}
\chi^{0}_{\gamma1}
&{}=
\ket{[(q_{1}q_{2})_{1}q_{3}]_{3/2}[(q_{4}q_{5})_{1}q_{6}]_{3/2}}_{0}\,,
\notag\\
%%%
\chi^{0}_{\gamma2}
&{}=
\ket{[(q_{1}q_{2})_{1}q_{3}]_{1/2}[(q_{4}q_{5})_{1}q_{6}]_{1/2}}_{0}\,,
\notag\\
%%%
\chi^{0}_{\gamma3}
&{}=
\ket{[(q_{1}q_{2})_{1}q_{3}]_{1/2}[(q_{4}q_{5})_{0}q_{6}]_{1/2}}_{0}\,,
\notag\\
%%%
\chi^{0}_{\gamma4}
&{}=
\ket{[(q_{1}q_{2})_{0}q_{3}]_{1/2}[(q_{4}q_{5})_{1}q_{6}]_{1/2}}_{0}\,,
\notag\\
%%%
\chi^{0}_{\gamma5}
&{}=
\ket{[(q_{1}q_{2})_{0}q_{3}]_{1/2}[(q_{4}q_{5})_{0}q_{6}]_{1/2}}_{0}\,,
\end{align}
\item $J=1$:
\begin{align}
\chi^{1}_{\gamma1}
&{}=
\ket{[(q_{1}q_{2})_{1}q_{3}]_{3/2}[(q_{4}q_{5})_{1}q_{6}]_{3/2}}_{1}\,,
\notag\\
%%%
\chi^{1}_{\gamma2}
&{}=
\ket{[(q_{1}q_{2})_{1}q_{3}]_{3/2}[(q_{4}q_{5})_{1}q_{6}]_{1/2}}_{1}\,,
\notag\\
%%%
\chi^{1}_{\gamma3}
&{}=
\ket{[(q_{1}q_{2})_{1}q_{3}]_{3/2}[(q_{4}q_{5})_{0}q_{6}]_{1/2}}_{1}\,,
\notag\\
%%%
\chi^{1}_{\gamma4}
&{}=
\ket{[(q_{1}q_{2})_{1}q_{3}]_{1/2}[(q_{4}q_{5})_{1}q_{6}]_{3/2}}_{1}\,,
\notag\\
%%%
\chi^{1}_{\gamma5}
&{}=
\ket{[(q_{1}q_{2})_{0}q_{3}]_{1/2}[(q_{4}q_{5})_{1}q_{6}]_{3/2}}_{1}\,,
\notag\\
%%%
\chi^{1}_{\gamma6}
&{}=
\ket{[(q_{1}q_{2})_{1}q_{3}]_{1/2}[(q_{4}q_{5})_{1}q_{6}]_{1/2}}_{1}\,,
\notag\\
%%%
\chi^{1}_{\gamma7}
&{}=
\ket{[(q_{1}q_{2})_{1}q_{3}]_{1/2}[(q_{4}q_{5})_{0}q_{6}]_{1/2}}_{1}\,,
\notag\\
%%%
\chi^{1}_{\gamma8}
&{}=
\ket{[(q_{1}q_{2})_{0}q_{3}]_{1/2}[(q_{4}q_{5})_{1}q_{6}]_{1/2}}_{1}\,,
\notag\\
%%%
\chi^{1}_{\gamma9}
&{}=
\ket{[(q_{1}q_{2})_{0}q_{3}]_{1/2}[(q_{4}q_{5})_{0}q_{6}]_{1/2}}_{1}\,,
\end{align}
\item $J=2$:
\begin{align}
\chi^{2}_{\gamma1}
&{}=
\ket{[(q_{1}q_{2})_{1}q_{3}]_{3/2}[(q_{4}q_{5})_{1}q_{6}]_{3/2}}_{2}\,,
\notag\\
%%%
\chi^{2}_{\gamma2}
&{}=
\ket{[(q_{1}q_{2})_{1}q_{3}]_{3/2}[(q_{4}q_{5})_{1}q_{6}]_{1/2}}_{2}\,,
\notag\\
%%%
\chi^{2}_{\gamma3}
&{}=
\ket{[(q_{1}q_{2})_{1}q_{3}]_{3/2}[(q_{4}q_{5})_{0}q_{6}]_{1/2}}_{2}\,,
\notag\\
%%%
\chi^{2}_{\gamma4}
&{}=
\ket{[(q_{1}q_{2})_{1}q_{3}]_{1/2}[(q_{4}q_{5})_{1}q_{6}]_{3/2}}_{2}\,,
\notag\\
%%%
\chi^{2}_{\gamma5}
&{}=
\ket{[(q_{1}q_{2})_{0}q_{3}]_{1/2}[(q_{4}q_{5})_{1}q_{6}]_{3/2}}_{2}\,,
\end{align}
\item $J=3$:
\begin{equation}
\chi^{3}_{\gamma1}
=
\ket{[(q_{1}q_{2})_{1}q_{3}]_{3/2}[(q_{4}q_{5})_{1}q_{6}]_{3/2}}_{3}\,.
\end{equation}
\end{enumerate}
Then the possible total wave functions for the $c^{3}b^{3}$ system are
\begin{enumerate}
\setcounter{enumi}{3}
\item Type D: $\varphi_{\text{D}}=\{c^{3}b^{3}\}$
\begin{enumerate}
\item $J=0$:
\begin{align}
\Psi_{\text{D}1}^{0}
={}&
\varphi_{\text{D}}
\otimes
\frac{1}{2}
\left(
\phi_{\gamma1}
\chi^{0}_{\gamma5}
-
\phi_{\gamma2}
\chi^{0}_{\gamma4}
-
\phi_{\gamma3}
\chi^{0}_{\gamma3}
+
\phi_{\gamma4}
\chi^{0}_{\gamma2}
\right)\,,
\notag\\
%%%
\Psi_{\text{D}2}^{0}
={}&
\varphi_{\text{D}}
\otimes
\phi_{\gamma5}
\chi^{0}_{\gamma1}\,,
\end{align}
\item $J=1$:
\begin{align}
\Psi_{\text{D}1}^{1}
={}&
\varphi_{\text{D}}
\otimes
\frac{1}{2}
\left(
\phi_{\gamma1}
\chi^{1}_{\gamma9}
-
\phi_{\gamma2}
\chi^{1}_{\gamma8}
-
\phi_{\gamma3}
\chi^{1}_{\gamma7}
+
\phi_{\gamma4}
\chi^{1}_{\gamma6}
\right)\,,
\notag\\
%%%
\Psi_{\text{D}2}^{1}
={}&
\varphi_{\text{D}}
\otimes
\phi_{\gamma5}
\chi^{1}_{\gamma1}\,,
\end{align}
\item $J=2$:
\begin{equation}
\Psi_{\text{D}1}^{2}
=
\varphi_{\text{D}}
\otimes
\phi_{\gamma5}
\chi^{2}_{\gamma1}\,,
\end{equation}
\item $J=3$:
\begin{equation}
\Psi_{\text{D}1}^{3}
=
\varphi_{\text{D}}
\otimes
\phi_{\gamma5}
\chi^{3}_{\gamma1}\,.
\end{equation}
\end{enumerate}
\end{enumerate}
%

%%%%%%%%%%%%%%%%%%%%%%%%%%%%%%%%%%%%%%%%%%%%%%%%%%%%%%%%%%%%%%%%%%%%
\section{The totally anti-symmetric wave functions for the $q^{N}$ clusters}
\label{App:wavefunc:QN}
%%%%%%%%%%%%%%%%%%%%%%%%%%%%%%%%%%%%%%%%%%%%%%%%%%%%%%%%%%%%%%%%%%%%

There are two methods to obtain the totally anti-symmetric wave functions for the $q^{N}$ clusters.
The first one is by using the properties of permutation group, namely the Clebsch-Gordon (CG) coefficients of the $S_{N}$ group.
The details can be found in Ref.~\cite{Stancu:1999qr}.
On the other hand, we can also obtain the wave functions by applying the quark exchange operators and imposing the anti-symmetric properties over the identical quarks [or $u$ and $d$ quarks (collectively as $n$) in the isospin $\mathrm{SU}(2)$ symmetry].
More precisely, assuming $\{\psi_{i},i=1,2,\ldots,n\}$ the total wave function bases.
The wave function that satisfies the Pauli principle can be expanded as their superposition
\begin{equation}
\Psi=\sum_{i}k_{i}\psi_{i}\,.
\end{equation}
The Pauli principle gives the secular equations
\begin{equation}
\sum_{j}\Braket{\psi_{i}|A|\psi_{j}}k_{j}=-k_{i}\,,
\end{equation}
where $A=\{A_{\alpha\beta}\}=\{A_{12},A_{13},\ldots\}$ are the operators that exchange the positions of $q_{\alpha}$ and $q_{\beta}$ in the wave functions.
Imposing these constraints and the orthonormal condition, we can obtain all possible wave functions that satisfy the Pauli principle.
As an example, here we use a simple system, the nucleon, to illustrate the methods.
First we use the method of the permutation group.
The total wave function is the direct product of the spatial, flavor, color and spin wave functions.
Note that for the nucleon, the spatial wave function ($R$) is totally symmetric, and the color wave function is totally anti-symmetric
\begin{equation}
\Yvcentermath1
\Psi
=
R
\otimes
\young(1,2,3)_{c}
\otimes
\young(123)_{sf}\,.
\end{equation}
For $I=S=1/2$, we have
\begin{align}
\Yvcentermath1
\young(123)_{sf}
%%%
={}&
\Yvcentermath1
S\left(\young(12,3),\young(12,3)\bigg|\young(123)\right)
\young(12,3)_{s}
\otimes
\young(12,3)_{f}
+
S\left(\young(13,2),\young(13,2)\bigg|\young(123)\right)
\young(13,2)_{s}
\otimes
\young(13,2)_{f}
\notag\\
%%%
={}&
\Yvcentermath1
\frac{1}{\sqrt{2}}
\young(12,3)_{s}
\otimes
\young(12,3)_{f}
+
\frac{1}{\sqrt{2}}
\young(13,2)_{s}
\otimes
\young(13,2)_{f}\,,
\end{align}
where
\begin{align}
\Yvcentermath1
S\left(\young(12,3),\young(12,3)\bigg|\young(123)\right)\,,
%%%
\qquad
%%%
\Yvcentermath1
S\left(\young(13,2),\young(13,2)\bigg|\young(123)\right)
\end{align}
are CG coefficients of the $S_{3}$ group, whose values can be found in Ref.~\cite{Stancu:1999qr}.
Next we use the exchange operators.
The matrix element $\Braket{\psi_{i}|A|\psi_{j}}$ can be viewed as a direct product of the matrix elements in the spatial ($R$), flavor ($F$), color ($\varphi$) and spin ($\chi$) wave function bases.
For the present case, we have
\begin{equation}
\Braket{R|A_{\alpha\beta}|R}=I_{1\times1}
\end{equation}
and
\begin{equation}
\Braket{\varphi|A_{\alpha\beta}|\varphi}=-I_{1\times1}\,.
\end{equation}
There are two possible spin wave functions
\begin{equation}
\chi_{1}=\Ket{\left(n_{1}n_{2}\right)_{1_{s}}n_{3}}_{1/2_{s}}\,,
\qquad
\chi_{2}=\Ket{\left(n_{1}n_{2}\right)_{0_{s}}n_{3}}_{1/2_{s}}\,.
\end{equation}
In these bases, we have ($A_{13}=A_{12}A_{23}A_{12}$ is not independent, and thus cannot give additional constraint for the wave function)~\cite{Park:2015nha}
\begin{equation}
A_{12}=
\begin{pmatrix}
+1\\
&-1
\end{pmatrix}\,,
%%%
\qquad
%%%
A_{23}=
\begin{pmatrix}
-\frac{1}{2}&\frac{\sqrt{3}}{2}\\
\frac{\sqrt{3}}{2}&\frac{1}{2}
\end{pmatrix}\,.
\end{equation}
Similarly, there are two possible flavor wave functions
\begin{equation}
F_{1}=\Ket{\left(n_{1}n_{2}\right)_{1_{i}}n_{3}}_{1/2_{i}}\,,
\qquad
F_{2}=\Ket{\left(n_{1}n_{2}\right)_{0_{i}}n_{3}}_{1/2_{i}}\,.
\end{equation}
In the space of the total wave function bases, $R\varphi\otimes\{\chi_{1}F_{1},\chi_{1}F_{2},\chi_{2}F_{1},\chi_{2}F_{2}\}$, the secular equations give
\begin{equation}
\begin{pmatrix}
-1\\
&+1\\
&&+1\\
&&&-1
\end{pmatrix}
\begin{pmatrix}
k_{1}\\
k_{2}\\
k_{3}\\
k_{4}
\end{pmatrix}
=
-
\begin{pmatrix}
k_{1}\\
k_{2}\\
k_{3}\\
k_{4}
\end{pmatrix}\,,
%%%
\qquad
%%%
\frac{1}{4}
\begin{pmatrix}
-1&\sqrt{3}&\sqrt{3}&-3\\
\sqrt{3}&1&-3&-\sqrt{3}\\
\sqrt{3}&-3&1&-\sqrt{3}\\
-3&-\sqrt{3}&-\sqrt{3}&-1
\end{pmatrix}
\begin{pmatrix}
k_{1}\\
k_{2}\\
k_{3}\\
k_{4}
\end{pmatrix}
=
-
\begin{pmatrix}
k_{1}\\
k_{2}\\
k_{3}\\
k_{4}
\end{pmatrix}
\end{equation}
or
\begin{equation}
k_{1}=k_{4}\,,
\qquad
k_{2}=k_{3}=0\,.
\end{equation}
Applying the normalization condition and choosing a suitable phase, we have
\begin{equation}
\Psi
=
\frac{1}{\sqrt{2}}
R\varphi
\left(
\chi_{1}F_{1}
+
\chi_{2}F_{2}
\right)\,.
\end{equation}
As expected, the two methods give same result.

In the following, we list the totally anti-symmetric wave functions for the $Q^{N}$ clusters with $N=3$, $4$, $5$, $6$.
\begin{enumerate}
\item The $Q^{3}$ cluster
\begin{enumerate}
\item $(\lambda\mu)=(11)$, $S=1/2$
\begin{equation}
\Yvcentermath1
\Psi_{1}^{N=3}
%%%
=
\frac{1}{\sqrt{2}}
\left(
\young(12,3)_{c}
\otimes
\young(13,2)_{s}
-
\young(13,2)_{c}
\otimes
\young(12,3)_{s}
\right)
\otimes
\young(123)_{f}\,,
\end{equation}
\item $(\lambda\mu)=(00)$, $S=3/2$
\begin{equation}
\Yvcentermath1
\Psi_{2}^{N=3}
%%%
=
\young(1,2,3)_{c}
\otimes
\young(123)_{s}
\otimes
\young(123)_{f}\,.
\end{equation}
\end{enumerate}
\item The $Q^{4}$ cluster
\begin{enumerate}
\item $(\lambda\mu)=(02)$, $S=0$
\begin{equation}
\Yvcentermath1
\Psi_{1}^{N=4}
%%%
=
\frac{1}{\sqrt{2}}
\left(
\young(12,34)_{c}
\otimes
\young(13,24)_{s}
-
\young(13,24)_{c}
\otimes
\young(12,34)_{s}
\right)
\otimes
\young(1234)_{f}\,,
\end{equation}
\item $(\lambda\mu)=(10)$, $S=1$
\begin{equation}
\Yvcentermath1
\Psi_{2}^{N=4}
%%%
=
\frac{1}{\sqrt{3}}
\left(
\young(12,3,4)_{c}
\otimes
\young(134,2)_{s}
-
\young(13,2,4)_{c}
\otimes
\young(124,3)_{s}
+
\young(14,2,3)_{c}
\otimes
\young(123,4)_{s}
\right)
\otimes
\young(1234)_{f}\,.
\end{equation}
\end{enumerate}
\item The $Q^{5}$ cluster
\begin{enumerate}
\item $(\lambda\mu)=(01)$, $S=1/2$
\begin{align}
\Yvcentermath1
\Psi^{N=5}
%%%
={}&
\Yvcentermath1
\frac{1}{\sqrt{5}}
\Bigg(
\young(12,34,5)_{c}
\otimes
\young(135,24)_{s}
-
\young(13,24,5)_{c}
\otimes
\young(125,34)_{s}
-
\young(12,35,4)_{c}
\otimes
\young(134,25)_{s}
\notag\\
&\Yvcentermath1
\qquad
+
\young(13,25,4)_{c}
\otimes
\young(124,35)_{s}
-
\young(14,25,3)_{c}
\otimes
\young(123,45)_{s}
\Bigg)
\otimes
\young(12345)_{f}\,.
\end{align}
\end{enumerate}
\item The $Q^{6}$ cluster
\begin{enumerate}
\item $(\lambda\mu)=(00)$, $S=0$
\begin{align}
\Yvcentermath1
\Psi^{N=6}
%%%
={}&
\Yvcentermath1
\frac{1}{\sqrt{5}}
\Bigg(
\young(12,34,56)_{c}
\otimes
\young(135,246)_{s}
-
\young(13,24,56)_{c}
\otimes
\young(125,346)_{s}
-
\young(12,35,46)_{c}
\otimes
\young(134,256)_{s}
\notag\\
&\Yvcentermath1
\qquad
+
\young(13,25,46)_{c}
\otimes
\young(124,356)_{s}
-
\young(14,25,36)_{c}
\otimes
\young(123,456)_{s}
\Bigg)
\otimes
\young(123456)_{f}\,.
\end{align}
\end{enumerate}
\end{enumerate}
%

%%%%%%%%%%%%%%%%%%%%%%%%%%%%%%%%%%%%%%%%%%%%%%%%%%%%%%%%%%%%%%%%%%%%
\section{Some formulae for the colorelectric and chromomagnetic interactions}
\label{App:formula}
%%%%%%%%%%%%%%%%%%%%%%%%%%%%%%%%%%%%%%%%%%%%%%%%%%%%%%%%%%%%%%%%%%%%

In this section, we introduce the $\text{SU}(6)_{cs}=\text{SU}(3)_{c}{\otimes}\text{SU}(2)_{s}$ group, then we can simplify the colorelectric and chromomagnetic interactions for hadron consists of at most \emph{two} flavors.

Following Ref.~\cite{Jaffe:1976ih}, the generators of $\text{SU}(6)_{cs}$ are defined by the products of $\sigma^{k}$ and $\lambda^{a}$
\begin{equation}
\alpha^{\mu}
%%%
={}
\left\{
\begin{split}
& \sqrt{\frac{2}{3}}\sigma^{k}, \quad k=1,2,3\,, \\
& \lambda^{a}, \quad a=1,2,\cdots,8\,, \\
& \sigma^{k}\lambda^{a}\,,
\end{split}
\right.
\end{equation}
which are normalized to $\text{tr}\alpha^2=4$.
The Casimir operators of the $\text{SU}(2)_{s}$, $\text{SU}(3)_{c}$ and $\text{SU}(6)_{cs}$ groups are defined as
\begin{align}
\text{C}_{2}
%%%
={}&
\bm{\sigma}\cdot\bm{\sigma}
=
4S(S+1)\,,\\
%%%
\text{C}_{3}
%%%
={}&
\bm{\lambda}\cdot\bm{\lambda}\,,\\
%%%
\text{C}_{6}
%%%
={}&
\sum_{\mu=1}^{35}
\left(\alpha^{\mu}\right)^{2}
%%%
=
\frac{2}{3}
\text{C}_{2}
+
\text{C}_{3}
+
\bm{\sigma}\cdot\bm{\sigma}
\bm{\lambda}\cdot\bm{\lambda}\,.
\end{align}
Now we can deal with the colorelectric and chromomagnetic interactions.

For $q^{s}Q^{t}$ system,
\begin{align}
&
\sum_{i<j}
m_{ij}
\bm{F}_{i}\cdot\bm{F}_{j}
\notag\\
%%%
={}&
m_{qq}
\sum_{i<i'=1}^{s}
\bm{F}_{i}\cdot\bm{F}_{i'}
+
m_{QQ}
\sum_{j<j'=s+1}^{s+t}
\bm{F}_{j}\cdot\bm{F}_{j'}
+
m_{qQ}
\sum_{i=1}^{s}
\sum_{j=s+1}^{s+t}
\bm{F}_{i}\cdot\bm{F}_{j}
\notag\\
%%%
={}&
\frac{m_{qQ}}{8}
\text{C}_{3}\left(q^{s}Q^{t}\right)
+
\frac{m_{qq}-m_{qQ}}{8}
\text{C}_{3}\left(q^{s}\right)
+
\frac{m_{QQ}-m_{qQ}}{8}
\text{C}_{3}\left(Q^{t}\right)
-
\frac{4}{3}
\times
\frac{sm_{qq}+tm_{QQ}}{2}\,.
\end{align}
For color-singlet hadron, the first term vanishes.
The colorelectric interaction becomes
\begin{equation}
\Braket{-\frac{3}{4}\sum_{i<j}m_{ij}\bm{F}_{i}\cdot\bm{F}_{j}}
%%%
=
\frac{sm_{qq}+tm_{QQ}}{2}
-
\frac{3}{8}
{\delta}m_{qQ}
\Braket{\text{C}_{3}\left(q^{s}\right)}\,,
\end{equation}
where
\begin{equation}
{\delta}m_{qQ}
=
\frac{m_{qq}+m_{QQ}-2m_{qQ}}{4}\,.
\end{equation}
As shown in Appendix~\ref{App:WaveFunc}, The bases have definite $\text{SU}(3)_{c}$ color representations for $q^{s}$ and $Q^{t}$ clusters.
Thus the colorelectric interaction is diagonal.
If $t=0$ or $t=1$, the formula can be further simplified, as shown in Eq.~\eqref{eqn:ECM:6Q} and Eq.~\eqref{eqn:Hcolor:cccccb}.

Similarly, the chromomagnetic interaction
\begin{align}
&
-\sum_{i<j}v_{ij}\bm{F}_{i}\cdot\bm{F}_{j}\bm{S}_{i}\cdot\bm{S}_{j}
\notag\\
%%%
={}&
-v_{qq}
\sum_{i<i'=1}^{s}
\bm{F}_{i}\cdot\bm{F}_{i'}
\bm{S}_{i}\cdot\bm{S}_{i'}
-v_{QQ}
\sum_{j<j'=s+1}^{s+t}
\bm{F}_{j}\cdot\bm{F}_{j'}
\bm{S}_{j}\cdot\bm{S}_{j'}
-v_{qQ}
\sum_{i=1}^{s}
\sum_{j=s+1}^{s+t}
\bm{F}_{i}\cdot\bm{F}_{j}
\bm{S}_{i}\cdot\bm{S}_{j}
\notag\\
%%%
={}&
-
\frac{v_{qQ}}{32}
\left(\text{C}_{6}-\frac{2}{3}\text{C}_{2}-\text{C}_{3}\right)_{q^{s}Q^{t}}
-
\frac{v_{qq}-v_{qQ}}{32}
\left(\text{C}_{6}-\frac{2}{3}\text{C}_{2}-\text{C}_{3}\right)_{q^{s}}
-
\frac{v_{QQ}-v_{qQ}}{32}
\left(\text{C}_{6}-\frac{2}{3}\text{C}_{2}-\text{C}_{3}\right)_{Q^{t}}
+
\frac{sv_{qq}+tv_{QQ}}{2}\,.
\end{align}
As shown in Appendices~\ref{App:WaveFunc}--\ref{App:wavefunc:QN}, the bases have definite $\text{SU}_{cs}(6)$ representations for the $q^{s}$ and $Q^{t}$ clusters, but not for the $q^{s}Q^{t}$ system.
Thus the bases will be mixed by the chromomagnetic interaction.
However, for $t=0$ and $t=1$ cases, the $q^{s}Q^{t}$ systems do have definite $\text{SU}_{cs}(6)$ representations [see Eq.~\eqref{eqn:rep:SU6:cccccb}], then the formula can be simplified to Eq.~\eqref{eqn:ECM:6Q} and Eq.~\eqref{eqn:HCM:cccccb:2}, respectively.
Note that similar formulae have also been obtained in Refs.~\cite{Jaffe:1976ih,Park:2015nha}.
%

%%%%%%%%%%%%%%%%%%%%%%%%%%%%%%%%%%%%%%%%%%%%%%%%%%%%%%%%%%%%%%%%%%%%
\end{appendix}
%%%%%%%%%%%%%%%%%%%%%%%%%%%%%%%%%%%%%%%%%%%%%%%%%%%%%%%%%%%%%%%%%%%%
%\bibliography{../../../my_reference/myJabref}
\bibliography{myreference}
%%%%%%%%%%%%%%%%%%%%%%%%%%%%%%%%%%%%%%%%%%%%%%%%%%%%%%%%%%%%%%%%%%%%
\end{document}